\documentclass[11pt]{article} % onecolumn, 11pt font
\usepackage[a4paper, margin=1.25in]{geometry} % A4 with 1.25 inch margins
\usepackage[utf8]{inputenc}

\usepackage[english]{babel}

\usepackage[normalem]{ulem}

\usepackage{subfig}

\usepackage{multirow} 
\usepackage{booktabs} % better lines

\usepackage{hyperref}
\usepackage{graphicx}
\usepackage{amsthm}
\usepackage{amssymb}

\usepackage[usenames,dvipsnames,svgnames,x11names]{xcolor}

\usepackage{algorithm}
\usepackage{algpseudocode}
\usepackage{amsmath}
\DeclareMathOperator*{\argmin}{argmin}

\usepackage{enumitem}

\usepackage{array}
\newcolumntype{L}[1]{>{\raggedright\let\newline\\\arraybackslash\hspace{0pt}}m{#1}}
\newcolumntype{C}[1]{>{\centering\let\newline\\\arraybackslash\hspace{0pt}}m{#1}}
\newcolumntype{R}[1]{>{\raggedleft\let\newline\\\arraybackslash\hspace{0pt}}m{#1}}

\usepackage{xspace}

\newcommand{\simtool}{\textsc{LoadStar}\xspace} 
\newcommand{\policy}{\textsc{Orbit}\xspace}
\newcommand{\forecast}{\textsc{Luna}\xspace}

\begin{document}

%% ARXIV STYLE
\title{\huge A Resource-centric Analysis and Optimization of NoSQL Workloads using Distressed Resource Volume Metric}
\author{ Gunika Verma$^{1}$, Aashutosh A V$^{1}$, Pooja Srinivas$^{1}$, Yogesh Simmhan$^{2}$, \\
Ayush Choure$^{1}$, Harshit Shah$^{3}$, Mayukh Das$^{1}$, Prashant Sasatte$^{3}$, \\
Chetan Bansal$^{4}$, Abhijit Pai$^{3}$, Suraj Dixit$^{3}$, Achint Agrawal$^{3}$
\thanks{To appear in Proceedings of the VLDB Endowment (PVLDB), Volume 19, Issue 9, 2026 (VLDB'26), \href{https://doi.org/10.14778/3819518.3819520}{doi:10.14778/3819518.3819520}.}
\\~\\ 
\textit{$^{1}$~M365 Research, Microsoft, Bangalore, India}\\ 
\textit{$^{2}$~Indian Institute of Science, Bangalore, India}\\ 
\textit{$^{3}$~Azure, Microsoft, Bangalore, India}\\ 
\textit{$^{4}$~M365 Research, Microsoft, Redmond, USA} 
\\~\\ 
\texttt{Email: \{mayukhdas, chetanb\}@microsoft.com},
\texttt{simmhan@iisc.ac.in} 
}
% %
\date{}

% make the title area
\maketitle

\begin{abstract}
Large-scale managed cloud databases leverage sophisticated load Packing and Migration (PAM) algorithms, which provide the efficiencies necessary for running these services at scale on cloud resources. 
Research into optimizing the resources and reliability of cloud databases at massive scales is limited by a lack of public NoSQL workloads.
We address this in the context of \textit{Cosmos DB}, 
Microsoft's flagship cloud-hosted NoSQL database.
We first propose \textit{open-source NoSQL workloads} from real Cosmos DB clusters, and analyze these traces to derive a novel reliability metric, \textit{Distressed Resource Volume (DRV)}, which captures the quality of service experienced by the end user.
We then develop an \textit{open-source policy simulation framework, \simtool,} powered by a non-parametric statistical model of estimating the QoS of real traffic patterns. These form a reusable benchmark pipeline for validating policies for resource-centric NoSQL workloads.
We then define a \textit{resource optimization problem} for placing Cosmos DB replicas onto VM nodes, develop the \forecast model for \textit{forecasting} future load distributions, and the \policy \textit{PAM algorithm} that uses these forecasts to trigger and rebalance stressed replicas, to reduce tail-errors.
Our experiments, validated using \simtool for these workloads, demonstrate \policy's benefits over the existing Cosmos DB policy and a worst-fit optimized baseline, with higher load delivered at lower error rates and up to $35\%$ reduction in resources. These have been deployed in production, with potential savings of $\$100M$s/yr while  improving service reliability for millions of customers.
\end{abstract}

\maketitle

\section{Introduction}

The last decade has witnessed the proliferation of varied \textit{NoSQL databases} to complement relational ones, including columnar data\-bases, key-value stores and graph databases~\cite{davoudian2018survey}. 
Besides alternative data models, these also leverage sharding and replication to achieve high scalability on commodity hardware while ensuring durability.
Alongside this has been the paradigm shift of business applications away from on-premise database deployments to large-scale \textit{cloud-hosted managed NoSQL databases} such as Cosmos DB and Dynamo DB from Cloud Service Providers (CSPs) such as Microsoft Azure and Amazon AWS.
This also shifts the responsibility for robust and efficient management of database operations to the CSPs.

The robustness guarantees of managed NoSQL databases have two competing objectives.
\textit{(1) Quality of Service (QoS) guarantees:} Ensuring high reliability and $6\sigma$ availability with strict adherence to Service Level Agreement (SLA) is essential to meet customer expectations. 
\textit{(2) Resource efficiency:} Reducing the operational costs (\textit{opex}, or Costs of Goods Sold, \textit{COGS}) is fundamental for lowering the price for clients while ensuring profitability for the CSP.
These are in opposition -- resource over-allocation helps meet the high SLA but leads to higher opex.
So, resource efficiency reduces to near-optimal packing and allocation of database entities onto compute resources, but is non-trivial in the presence of dynamic, sometimes unpredictable, load patterns that can impact reliability guarantees.

\textit{Resource management and load balancing} are critical functionalities of cloud infrastructure orchestration that impacts cloud databases, with diverse CSP platforms such as AWS's Aurora~\cite{barnhart2024resource}, Google's Borg~\cite{verma2015large} 
and Microsoft's Service Fabric~\cite{servFab18}.
Resource management is the central theme of enterprise relational databases such as BRAD~\cite{kraska2023check} and Moneyball~\cite{poppe2022moneyball}. Alibaba's resource management framework, \textit{Eigen}~\cite{eigen23}, optimizes utilization across active/warm/cold nodes for database workloads, with elements of forecasting and vector bin-packing on resource types. There is also much work on classical and cloud applications using vector bin-packing heuristics to solve this allocation problem~\cite{panigrahy2011heuristics,christensen2017approximation,wang2014dominant,carl-tcc-2025}. While bin-packing is NP-hard with poor approximations~\cite{BrandaoP16}, the ability to forecast the node-level loads can make them tractable.

We consider these in the context of  \textit{Azure Cosmos DB}, a cloud-hosted NoSQL database from Microsoft that supports unstructured, semi-structured, structured, and vector data types. It uses a partitioned and replicated deployment for each customer database, with \textit{replicas} placed on \textit{nodes} hosted on Azure VMs with multi-tenancy.

Load-balancing optimization problem formulations often treat QoS or reliability as a \textit{global constraint} in order to \textit{optimize efficiency}. ``Reliability'' here is a general catch-all concept capturing end-user experience including all types of SLA guarantees. However, a key user-facing metric that is overlooked is \textit{error rates} when performing database operations. These are affected by the stochastically varying demand and stress on the underlying resource caused by the replica placement. These include latency, throughput, availability, etc. errors. Just setting minimum reliability constraints is either inadequate or, often, counter-productive to achieve efficiency and low error rates since the compute load has a direct probabilistic relationship with the nett load. Assuming all load distributions are equal, setting a global reliability constraints (e.g., CPU usage $\leq 80\%$) does not eliminate \textit{tail errors}. For products with large underlying infrastructure and a need to operate at very high availability, \textit{these tail errors can add up to a failure event}.

It is paramount to design load optimizers that enhance usage efficiency (lower COGS) jointly with stochastic quantification of QoS for NoSQL workloads, subject to varying and uncertain load patterns. But key challenges in solving these are: (1) A dearth of principled and relevant real-world \textit{workloads and metrics to quantify QoS (reliability)} against varying demand and load; (2) The absence of a \textit{high-fidelity simulation framework} to emulate managed databases and rapidly validate strategies before deployment; and finally, (3) A lack of tractable, temporally-aware, adaptive \textit{load balancing strategies} that optimize resource efficiency, subject to such a reliability metric. This EAB article addresses these gaps.

In this paper, we make the following contributions:
\begin{enumerate}[leftmargin=0cm,itemindent=.5cm,labelwidth=\itemindent,topsep=0pt]
\item 
    We \textit{open-source benchmark traces for NoSQL workloads}, for three Cosmos DB clusters each having approximately 190 nodes and 32k-42K database replicas, reporting resource usage and errors over 5-6 weeks. We analyze this workload to develop a novel user-centric reliability metric, \textit{Distressed Resource Volume (DRV)}, that reflects the practical QoS experienced by end-users for the allocated compute capacity to their NoSQL database (Sec.~\ref{sec:workload}).

\item 
    We develop an open-source \textit{Policy Simulator framework, \simtool}, that uses non-parametric statistical models based on the workload traces, to accurately predict error rates with low variance/noise, for a given database workload and placement policy, and also models node-level errors to estimate DRV (Sec.~\ref{sec:simulation}).

\item     
    We define a resource optimization problem to place Cosmos DB shards and replicas onto VM nodes to improve the DRV metric.
    We design a \textit{resource placement algorithm, \policy}, to solve this  (Sec.~\ref{sec:packingpolicy}), by leveraging
    a non-parametric distributional \textit{forecast model, \forecast}, for replica level loads based on a systematic study of load distributions across the replica population (Sec.~\ref{sec:forecast}).

\item 
    We perform use our benchmark pipeline consisting of \simtool framework and Cosmos DB workloads to perform detailed experiments that compare our \policy policy against the Simulated Annealing policy used in Cosmos DB and a worst-fit baseline. We report the serving load at 
    different error rates, the resource load uniformity, and the migration counts (Sec.~\ref{sec:results}). All these favor \policy, and help it
    minimize resource usage by up to $35\%$, and hence costs, while keeping error rates within SLA.

\end{enumerate}
    The user-centric DRV metric, open-source NoSQL workloads, and the simulation framework form a reusable benchmark pipeline (\url{https://aka.ms/LoadStar}) that can help the community develop and validate new forecasting, packing and migration strategies for cloud-scale NoSQL databases, and practitioners validate their own workloads. Our proposed policies have also been deployed to production Cosmos DB, with \$100M+/yr savings.

\section{Workload Analysis and DRV Metric}
\label{sec:workload}

\begin{figure}[t!]
    \centering
    \includegraphics[width=1\columnwidth]{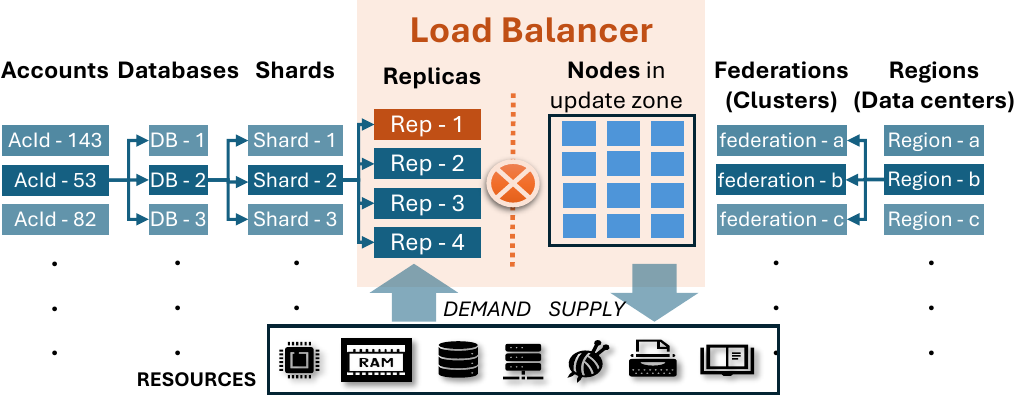}
    \caption{Cosmos DB resource provisioning hierarchy}
    \label{fig:systemArch}    
\end{figure}

Microsoft Cosmos DB~\cite{GuayPaz2018} is a managed NoSQL database hosted on Azure cloud. 
Cosmos DB is active in 60+ global data centers, and runs on $100,000$s of nodes which host tens of millions of database shard replicas. The annual hardware costs for tens of millions of allocated CPU cores is significant. So, any improvements to COGS and QoS for Cosmos DB benefits the substantial user base and costs.

\subsection{System Model of Azure Cosmos DB}

A Cosmos DB \textit{account} is an individual business client (Figure~\ref{fig:systemArch}). Each account has multiple \textit{databases}, each database has multiple collections, and every collection has several shards. \textit{Shards} are formed from database partitioning and every shard generally has 4 \textit{replicas}: one primary, which gets the first write and no read traffic, and three secondaries, which handle all read traffic and federated writes. Each replica is the unit of placement onto a node (VM).

Cosmos DB uses the Azure Service Fabric (SF)~\cite{servFab18} to manage Cosmos DB's 
infrastructure on 100,000s of \textit{nodes}.
A data center region has \textit{clusters}, which contain 100s of nodes each. 
A shard usually has all its replicas in the same cluster.
This streamlines the load balancing operations by limiting active load management and \textit{replica migrations} to primarily within a cluster.

Each node has an upper bound on the number of replicas it can host,
limited by a capacity constraint on CPU, Memory, Storage, I/O, Threads, Reads, and Writes. The individual resources are normalized into a standard measure, \textit{Requested Units (RU)}, which is the unit of provisioning, allocation and billing of resources. 
RU is a unitless compound resource usage metric used to bill users for Cosmos DB. Typically, 1 RU supports 1 key-value lookup per second~\cite{cosmos-cap-calc}.
A more realistic query such as \texttt{\textbf{SELECT VALUE SUM}(c.pricing.grandTot) \textbf{FROM} c \textbf{WHERE} c.createdAt >= '2025-08-25T00:00:00Z' \textbf{AND} c.createdAt < '2025-09-24T23:59:59Z'} takes $11.66$ RUs.

Each shard has a unique RU traffic profile. The replica allocation strategy determines if individual replicas will have enough resources. The general strategy to mitigate resource contention on a node is to redistribute some replicas to other nodes to achieve better load balancing.
But, this results in \textit{replica migrations}, which consumes resource bandwidth and takes the migrating replicas temporarily offline. 
This evokes three key questions that form the \textit{Packing and Migration (PAM) policy}:
\textit{who-to-migrate, when-to-migrate} and \textit{where-to-migrate}, and impacts overall delivery stress, migration volumes, disruptive migration count and compute availability.

\begin{table}[t!]
  \centering
  \caption{Workload Description of Different Clusters}\label{tab:data_desc_base}
  
  \small
\renewcommand{\arraystretch}{0.9} 
  \begin{tabular}{l|r|r|r|r}
    \toprule
    \textbf{Metric $\downarrow$~\qquad~Cluster $\rightarrow$} & \bf{C1a} & \bf{C1b} & \bf{C2} & \bf{C3} \\
    \midrule
    Time Duration (days) & 60 & 13 & 54 & 54\\ \hline
    \# of Nodes in Cluster & 196 & 196 & 196 & 190\\ 
    \# of Partitions & 7,953  & 8,798  & 7,960 & 11,593 \\
    \# of Replicas & $\approx 32k$  & 35,408  & 31,207 & 42,876 \\ 
    Max. Load on Cluster (RUs) & 3.7M & 3.7M & 4.3M & 8.5M \\
    Max. CPU usage (agg. over cores) & 1.83 & .68 & 1.48 & .79 \\
    Max. Mem. Usage (GB) & 
    97.0
    &
    15.3
    &
    98.2
    &
    107.1
    \\
    Max. Storage Usage (TB) & 
    2.60  
    &
    0.17
    &
    2.73
    & 
    2.89
    \\ 
    \bottomrule
  \end{tabular}
  \label{tab:dataset1}
  
\end{table}

\subsubsection{Simulated Annealing Policy}
\label{sec:sa}
Azure Service Fabric's Load Balancer's PAM policy uses \textit{Simulated Annealing (SA)}~\cite{simAnn1,vanLaarhoven1987,simAnn2}, a statistical physics-based optimization for NP-complete scenarios. It has two modes of triggering a re-balancing of replicas in a cluster: \textbf{(1)} Observing a \textit{constraint violation} on one node, or \textbf{(2)} Observing the max/min ``energy ratios'' of SA cross preset thresholds. Here \textit{energy} is a user-defined abstract function that captures the instability of the system state, e.g., the ratio of maximum to minimum node utilization in the cluster. The underlying resource metrics (e.g., CPU usage, vacant RAM), the constraints on these metrics, and the energy function are configured for specific deployments. When a re-balancing is triggered, the algorithm does a time-bound search of neighboring configurations and moves to the (nearly) best configuration it finds. This reassignment is applied over a set of replicas-to-node mappings, with 100s of migrations done in parallel.

\subsubsection{A major challenge?} 
Although SA policies are expected to be efficient and locally effective, our early analysis showed a significant imbalance in replica assignments, which negatively impacts cluster-level utilization and migration costs at a cloud-scale. While load balancing and distributed database optimizations are well-studied, existing solutions take a restricted view. Some solve the system-level resource management problem irrespective of how cloud database workloads behave, while others address aspects of distributed databases, but with constraints on stability and migration failures or an emphasis on query reliability~\cite{eigen-vldb-2023, accordion-vldb-2014}. 
We see a gap in \textit{fault-aware load packing and resource management} for managed NoSQL databases on the cloud that goes beyond reliability or cost constraints and jointly optimizes key dimensions of migration costs, workload performance, scalability and fault rates.

Key bottlenecks in achieving this are: (1) Lack of real-world open-source \textit{NoSQL workloads} and fault profiles of large-scale cloud-managed databases, (2) Lack of \textit{meaningful metrics} to jointly characterize relevant dimensions, and (3) Lack of a robust \textit{policy simulator} to evaluate large-scale load profiles and fault/error distributions before deployment.
The subsequent sections address these.

\begin{figure}[t!]

    \centering
      \subfloat[Cumulative Distribution across Replicas (all metrics) for C1b]{
        \includegraphics[width=0.49\columnwidth]{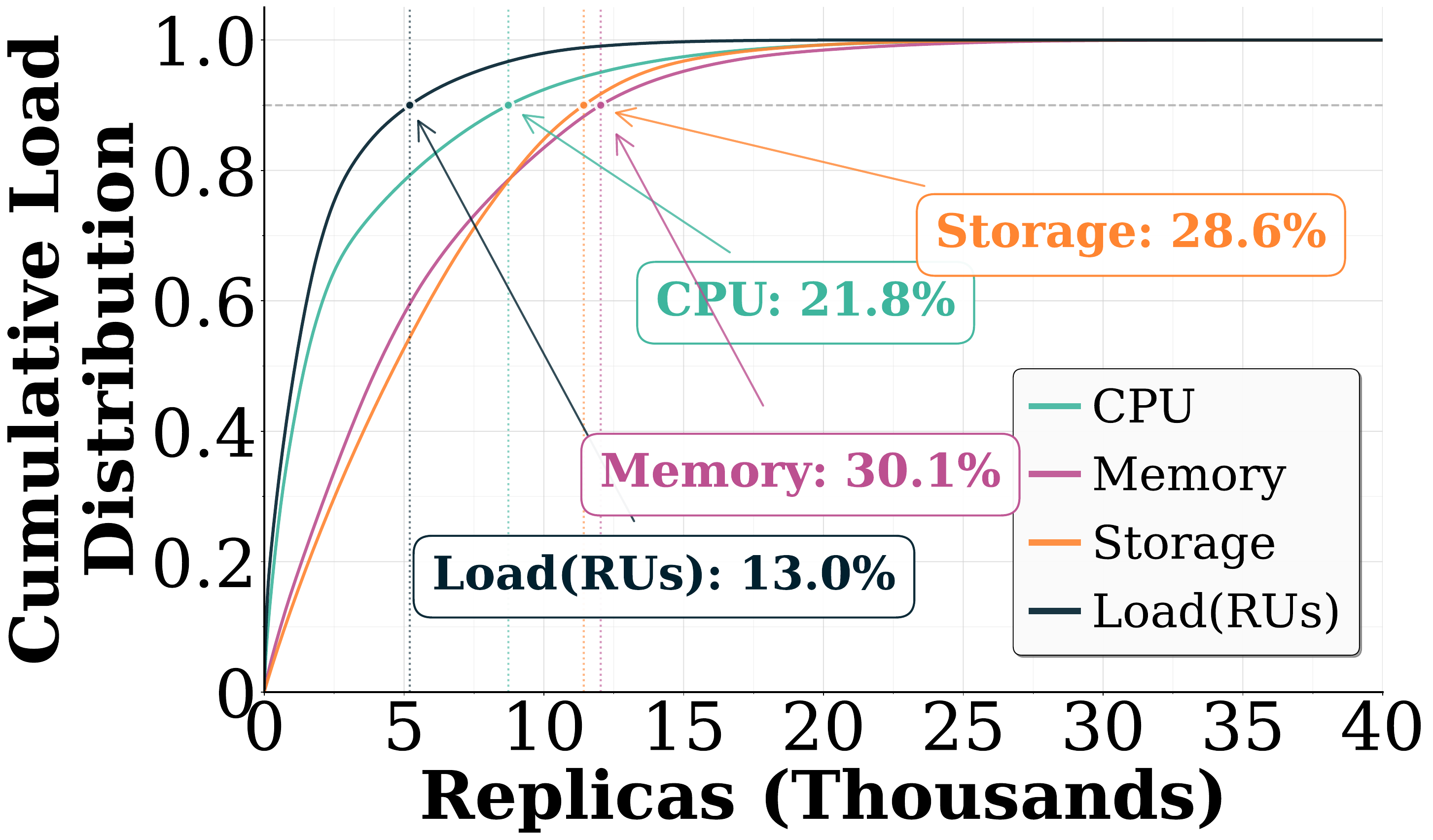}
        \label{fig:CumulativeStorage}
      }~
      \subfloat[Cumulative Distribution across partitions]{
        \includegraphics[width=0.49\columnwidth]{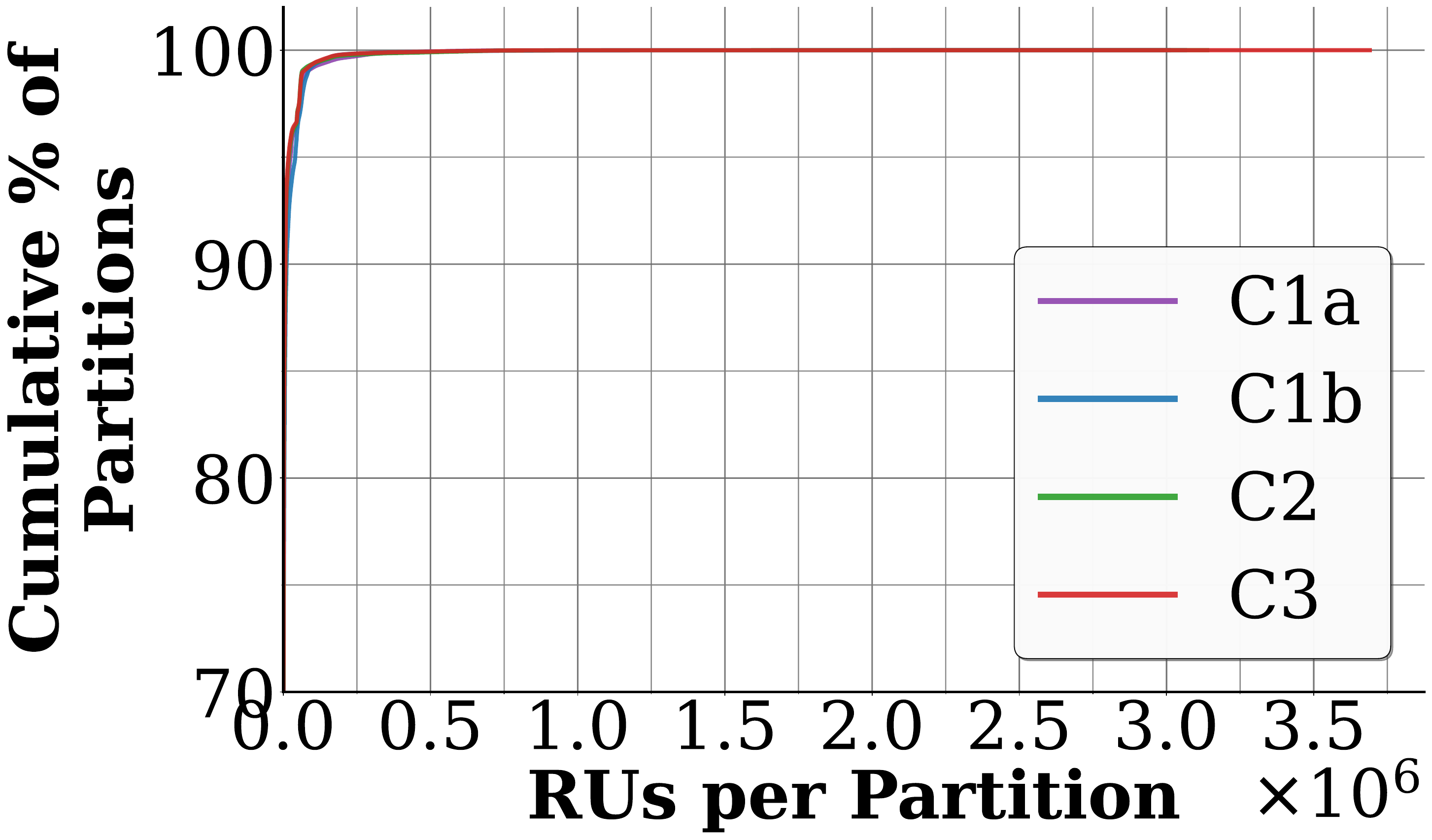}
        \label{fig:cdf_partition_vs_RUs}
      }
      
    \caption{Resource Usage Distribution}

\label{fig:CumulativeDistributions}

\end{figure}

\subsection{Cosmos DB NoSQL Workload Dataset}

We propose an open benchmark dataset based on real-world NoSQL production workloads from Cosmos DB, and use it for our experimental analysis. We extract, transform, and consolidate performance telemetry of Cosmos DB and the underlying normalized hardware for NoSQL requests with varying proportions of loads (RUs). These workloads are from diverse applications, such as LLM infrastructure, e-commerce websites, financial products, and enterprise productivity software deployed in North American data centers. They have been suitably anonymized and masked so as to not reveal any specific (de-anonymizable) details about the clients, topology, or fine-grained infrastructure.

The workloads are from \textit{three Cosmos DB clusters}, C1, C2 and C3, with over 190 VMs each hosting about $8,000$ shards, serving up to 8.5M RUs of load (Table \ref{tab:dataset1}). Of these, C1a and C1b come from the same C1 cluster but for different time periods, with the smaller dataset C1b better suited for computationally expensive ablation studies or creating train--test data separation. 
All the clusters have homogeneous nodes with identical resources. Individual nodes 
have a memory threshold of 380GB and storage of 5.2TB. 
While the CPU core usage ranges from $[0.0,1.0]$, we do not normalize across cores; hence, the utilization per node can add up to more than $1$ with the per core usage bounded at $0.6$.

\begin{figure}[t!]

    \centering
      \subfloat[Weak correlation between load \& error rates at lower loads, for nodes in Cluster C1b for a week in March 2025.]{
        \includegraphics[width=0.49\columnwidth]{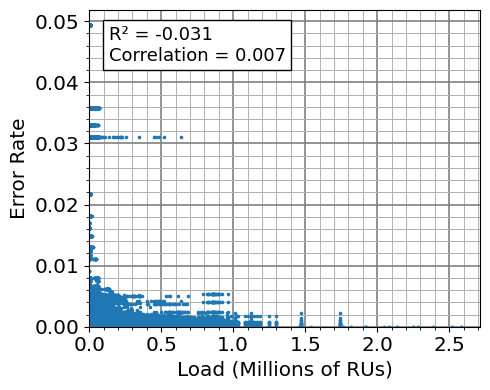}
        \label{fig:load-error-scatterplot}
      }~
      \subfloat[Error Rates increase super-linearly at higher load bins (tail).]{
    \includegraphics[width=0.5\columnwidth]{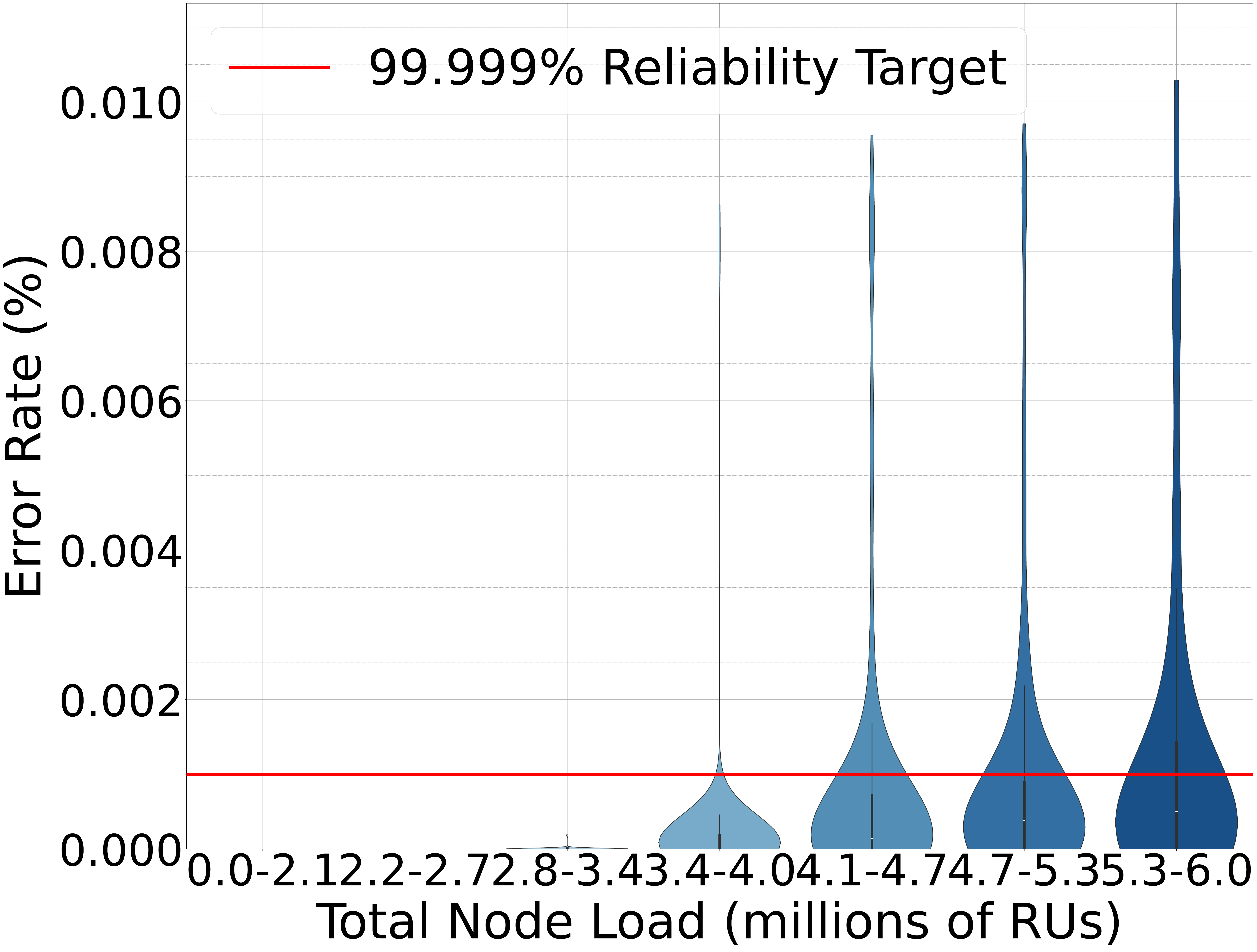}
    \label{fig:error-rates-load}
      }
   
    \caption{
    Non-linear correlation between load and error rate.
    }
    \label{fig:loaderrorrel}
   
\end{figure}

Figure~\ref{fig:CumulativeStorage} illustrates for \textbf{C1b} a Cumulative Distribution Function (CDF) plot of the four key resource dimensions: request load (RU), CPU, memory and storage, with increasing replica counts. Although correlated, differences in usage among these resources, especially at the \textit{tails} -- since the CDFs do not converge at the same point -- indicate unbalanced load packing. 
\\
The CDF of partition level load (RUs) in Figure~2b shows that clusters in general have low load per partition, but have a long tail (1 ppm) of higher load. Drilling in, the scatter plot in Figure~3a between the node loads and error rates shows a weak correlation for lower RUs, and is dominated by nodes with low load. So, tail errors merely manifest as borderline SLA violation. But, as we show 
in Figure~\ref{fig:error-rates-load}
loads of $\geq 3.0$ RUs cause a super-linear rise in error rates, which, while infrequent can violate Cosmos DB's guarantees of \textit{five-nines} ($99.999\%$) of availability. So, current workload packing policies 
purely based on resource constraints are inadequate. 
We need to jointly analyze the RU load and error rate distributions.

\subsection{Reliability and Efficiency Tradeoffs}

A key aspect of this article is to practically improve \textit{experienced reliability}, beyond mere SLA constraints. While the load optimization policy aims to optimize efficiency of compute allocation, the reliability experienced by end users determines if a particular replica-node assignment is of acceptable \textit{quality}. 
Cosmos DB has a target SLA of five-nines of availability. It has a comprehensive error tracking system that logs every low-level failure with specific \textit{error codes}. As replicas run out of resources, errors related to latency, throughput, availability, etc. start growing. At certain threshold, SLA monitors will report \textit{availability incidents}, requiring intervention.
Reliability and utilization efficiency form competing objectives here, which makes this more nuanced than a simple bin-repacking.

Figure~\ref{fig:error-rates-load} shows a distribution of error rates reported on replicas as the load increases. 
Here, we simulate loads of up to 6.3M RUs per node on a cluster of 213 nodes, 
and sample logs every 5~mins to record the error rates. The min-max RU load range are binned into seven equal-width intervals, 
except for the initial underflow bin. We resample the binned data using probabilistic bootstrapping and Kernel Density Estimation (KDE) to ensure all bins have the same sample size and obtain a continuous data density across the bins.

The adverse impact of load on error is clear.
At RU loads of up to 3.46M, the errors rates are low and well below the QoS threshold of 0.001\% (99.999\% reliability). Beyond this, however, the fourth bin has 1.2\% of samples exceeding 0.001\% error rate and last bin has 39.6\% of violations exceeding 0.001\%; errors above 0.01\% are treated as anomalies and censored. There is a clear monotonic decline in reliability.
While the throughput constraints are not yet being violated, the error rate 
is impacting the QoS for the packing policy used. 
Since errors climb non-linearly with load, a more uniform load distribution will always provide better reliability than a lop-sided ones. While this follows from Jensen's inequality, we later observe this in our empirical tests too.
These motivate the need for \textit{non-traditional user-facing metrics of reliability}, and specialized optimization strategies to address this at scale.

We capture the \textit{uniformity of load and utilization} using the \textit{\textbf{P80/\allowbreak P20} ratio}, which is the utilization values of the $80^{th}$ percentile node over the $20^{th}$ percentile node for a metric. In a perfect system with uniform load, $\frac{P80}{P20} \to 1.0$. Fig.~4 shows the $\ln{\frac{P80}{P20}}$ ratio of various resource metrics (CPU/RAM/Storage) for a week. They are unevenly utilized and the RU is even more skewed. So, \textit{even if the average load per node is low, a placement strategy that leads to uneven distribution can create QoS hotspots.}
Alternative ratios such as P90/P10 and P70/P30 also show similar trends.

Figure~\ref{fig:Cumulative_NodesvsRUs} shows the real distribution of RUs across nodes in our benchmark datasets, C1b, C2 and C3. \textit{All} of them have nodes that are serving $>3$M RUs, which can lead to policy violations as see in Figure~\ref{fig:error-rates-load}, while \textit{none} of them have very high average load density. Such a \textit{long-tail phenomena} leading to higher error rates can be mitigated only by controlling the utilization variance. Next, we propose a novel metric to capture this user-facing QoS.

\begin{figure}[t]

    \centering    \includegraphics[width=0.8\textwidth]{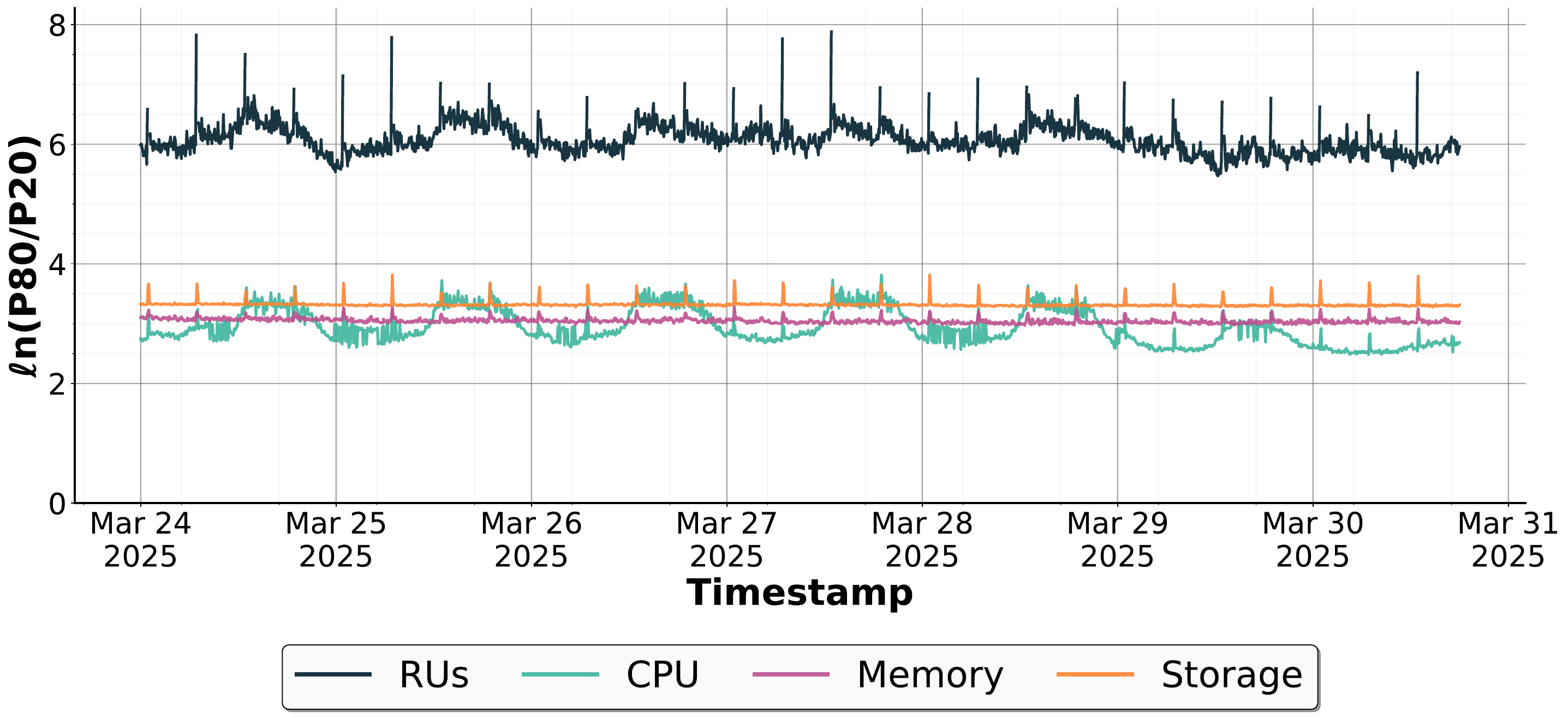}
    
    \caption{P80/P20 plots for all metrics for Cluster C1b.}
    \label{fig:Load-P80/p20-across-nodes-SA}
    
\end{figure}

\subsection{Distressed Resource Volume (DRV) Metric}
\label{sec:drv}

Cosmos DB maintains error rates for various hierarchical \textit{error codes} (or types) as
a time series of counts by error code at the replica or node level. They can be aggregated for a shard, node, account, etc.
\begin{figure}[t!]
    \centering
      \subfloat[Cumulative Percentage of Nodes as a function of number of RUs on clusters Baseline(C1b), C1b, C2, C3]{
        \includegraphics[width=0.49\columnwidth]{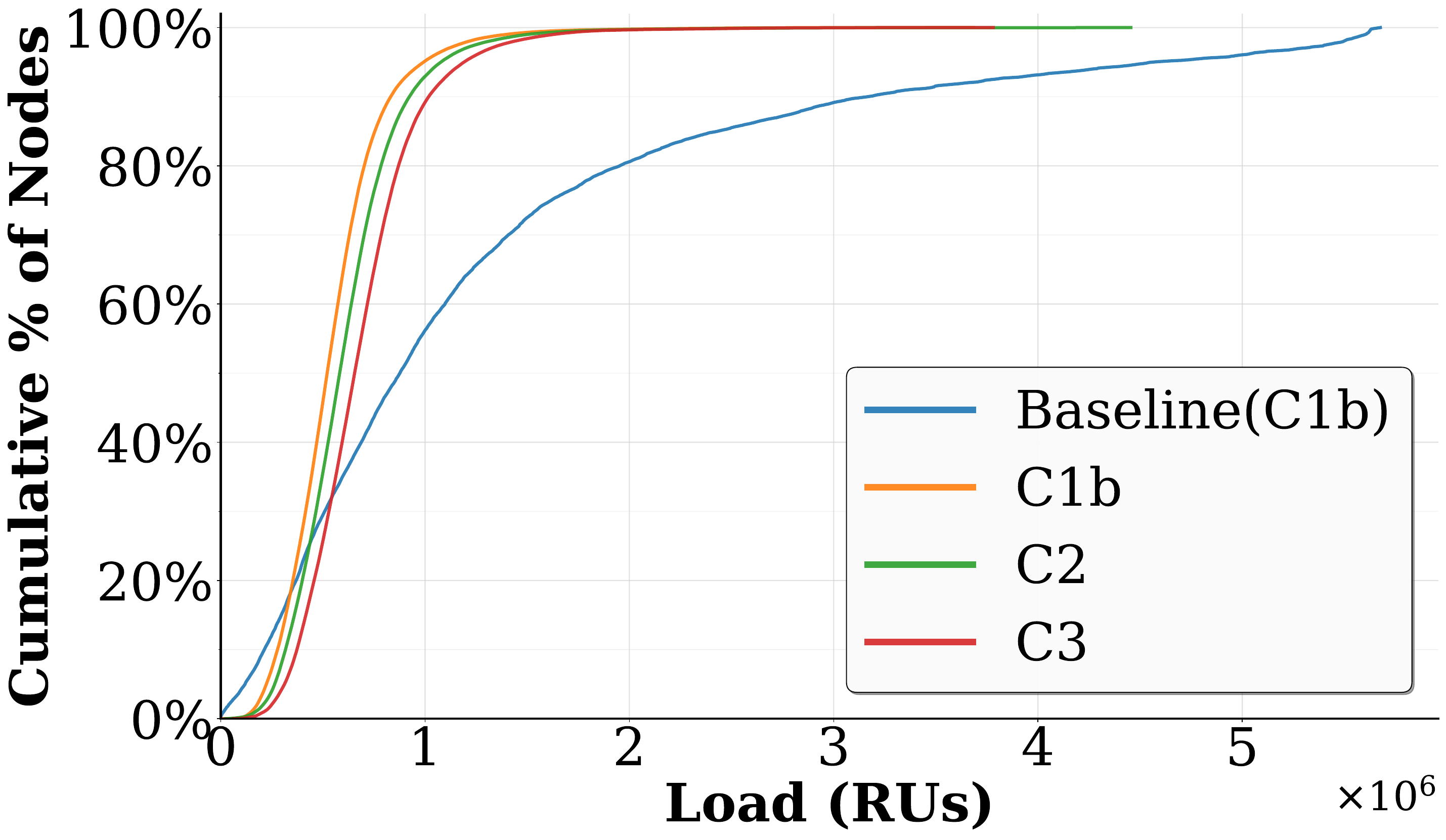}
        \label{fig:Cumulative_NodesvsRUs}
      }~
      \subfloat[Example DRV profiles for 2 migration policies 
    (\textit{ln} of requests)]{
        \includegraphics[width=0.49\columnwidth]{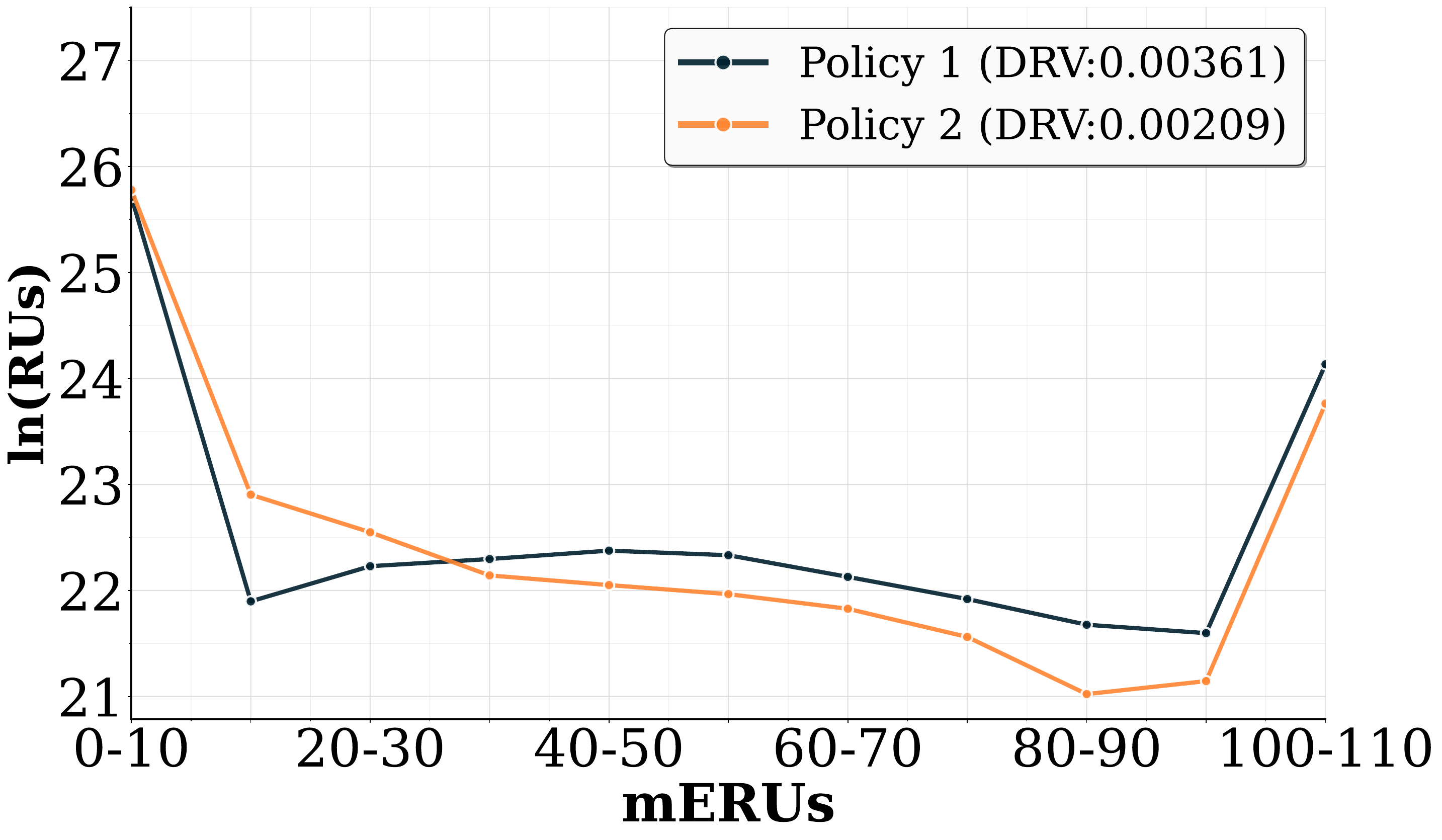}
        \label{fig:drvComparisonExampleTwoPolicies}
      }
      
    \caption{DRV profile of observed and predicted ERU values}
    \label{fig:drvProfile}
    
\end{figure}
A na\"ive reporting of total error counts over long time periods has no statistical merit and can be misinterpreted. The generation of an error code by itself may not be a concern. Although ideally no errors should occur, complex large-scale systems behave stochastically, even under no duress. At Cosmos DB's scale, with millions of partitions hosted on hundreds of thousands of nodes, some base emission of error codes will keep happening.

A clearer indicator of a fault is error codes emitted over longer durations that indicate a non-transient SLA-impacting scenario. The RUs delivered to customers during these periods are at a QoS inversely proportional to the number of errors in that period.
This leads to our novel proposed metric, 
\textit{Distressed Resource Volume (DRV)}, which reports the RUs (or any specific resource dimension, such as CPU cycles, RAM, storage, etc.) delivered at various levels of reliability (or conversely, distress).

Formally, let $\delta$ be a uniform time window (often a function of the sampling frequency of resource counters) within which we measure the volume of errors.
For a given \textit{node} $M^i$ at \textit{time} $t$, let $E^i([t - \delta, t])$ be the \textit{number of errors} across all types emitted by this node between times $t - \delta$ and $t$. With $\mathcal{R}(M)$ as the set of \textit{replicas} hosted on the node $M$, and $I_{t}$ being the \textit{resource consumption} of the specific replica, the RU-normalized error rate, called \textit{ERU (Errors per RU)}, for node $M$ at time $t$, is defined as:
\setlength\abovedisplayskip{2pt}
\setlength\belowdisplayskip{2pt}
\[
\text{ERU}(M, t) = \frac{E_M([t - \delta, t])}{\sum_{I_t \in \mathcal{R}(M)} I_t}
\]
This is essentially a measure of errors emitted \textit{per unit} of load delivered (RU). Using this, we split the entire range of ERUs into a fixed number of uniform error bins, $EB_{i} (i \leq d)$, with the last overflow bin being unbounded. We then define the \textit{DRV profile} as a $d$-dimensional vector, where the $k^{th}$ component is the amount of load delivered at the ERU value belonging to the $k^{th}$ bin. Formally,
\begin{equation}
    \text{DRV}_{k}(\mathcal{M}, \mathcal{I}, \Omega) = \sum_{t, R(I_t) \in \mathcal{R}(M):E(M,t)\in EB_{k}} I_t
\end{equation}
This novel profile vector captures the complete state of reliability and user-facing delivery quality.

\noindent\textit{Example}
Figure~\ref{fig:drvComparisonExampleTwoPolicies} compares the impact of two different replica placement policies on the overall reliability, based on their DRV profiles.
The Y-axis is the volume of compute delivered (RUs) 
and X-axis encodes the ERU bins. The area under this curve equals the total compute delivered. Different policies packing the same total RUs will have the same area under the curve but different slopes at various points. Ideally, the packing policy should have the most load delivered towards the left regions (low ERU and better reliability) and lower loads as we go right, having lower reliability. 
In this example, the orange policy in general delivers more RUs at a lower ERU rate compared to the blue policy.

\section{\simtool Policy Simulation Framework}
\label{sec:simulation}

\begin{figure}[t!]
        \centering    
\includegraphics[width=1\columnwidth] {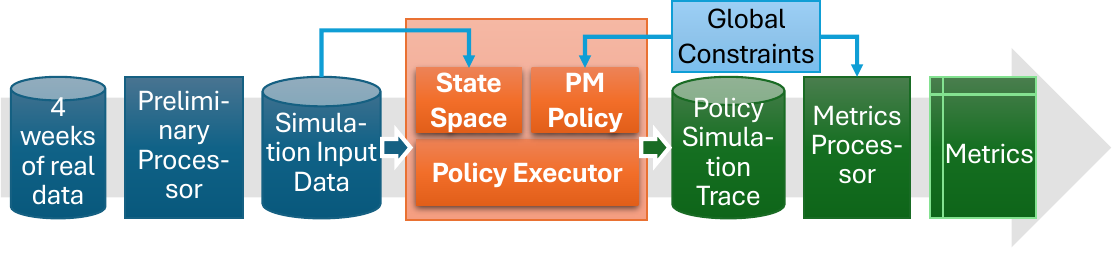}

        \caption{\simtool Policy Simulation Framework for Cosmos DB Load Balancer}
        \label{fig:testbedDesign}
        
\end{figure}

Before deploying any new resource management policies on massive-scale infrastructure, cloud providers validate their impact using policy simulators. This help uncover potential issues with poor performance, which can affect the experience of millions of users, or accrue tens millions of dollars of costs, before actual deployment.
Existing PAM policy simulators used in industry tend to be proprietary.
While they simulate the resource utilization behavior with the aim of minimizing constraint violations, they fail to provide insights into the experienced reliability of the end user.

We propose a new \textit{open-source PAM policy simulation framework, \simtool,} which simulates the load and corresponding error rates for Cosmos DB.
The key technical contribution is the \textit{non-parametric modeling} of ERU--RU relationship; we demonstrate empirically the effectiveness of these techniques in the context of Cosmos DB workloads. This framework easily scales to multiple clusters across multiple weeks.
The \simtool simulation framework is now \textit{actively used in Cosmos DB} to assess reliability profiles of various scenarios.
Later in this paper, we use it to estimate the ERUs that will be generated for replicas based on the node-level RUs achieved by our PAM policy for a given workload, i.e., go from estimated resource metrics to the estimated end-user QoS. 
\simtool is available as part of the public artifact for this article.

\subsection{Design}

Figure \ref{fig:testbedDesign} describes the overall schematic of \simtool with the following key steps and modules.
The \textit{Preliminary Processing} module converts the raw server logs for a past workload into a standard form, addressing source level idiosyncrasies, e.g., some hosting systems change the identifier of a replica every time it moves, and assigns unique IDs to each replica.
The resulting \textit{Simulation Input Data} is in a standardized form with just the necessary columns, viz., each replicas node's location at every timestamp, the utilization of all resources, and the errors being emitted.

The \textit{Policy Executor} generates the simulation trace for the \textit{new} policy encoded in the policy sub-module. It interfaces with several sub-modules,
which correspond to functions defined in the optimization problem formulation.
The \textit{State Space Manager} is a global configuration cache that tracks the current placement of all replicas.
The \textit{PM policy} module implements the PAM policy $\Omega(.)$ that is invoked along with contextual parameters, e.g., load (RU) forecasts, and generates a time ordered sequence of node--replica configurations that are emitted into the \textit{policy simulation trace}.

The \textit{Metrics Processor} module first enriches the generated policy trace with error rate estimates using the models discussed in Section~\ref{sec:errorGenModels}. The simulation trace is then converted to the same schema as the input simulation data. The metrics processor then generates the aggregate metrics (such as DRV) into a final metrics file, which can then be analyzed to understand the impact of the policy on the system behavior for the given workload.

This framework can also be used to simulate an \textit{alternate cluster configuration}, by updating the default global state space manager configuration used for simulation
with the updated configuration. We use this when reducing the node footprints in Section~\ref{sec:results:nodes}.

\subsection{Non-Parametric Models to Generate ERU Estimates for RU Levels}
\label{sec:errorGenModels}

We use non-parametric statistical models to estimate the ERU values for given RU levels. These do not make any distributional assumptions and are effective in real data situations. We describe the method, followed by convergence tests on the C1 datasets.

\subsubsection{Methodology}
We split the given RU range in the input trace into bins, and for each bin we resample to achieve the target size of ERU entries per bin. For this, we randomly generate RU values within the bin range and find the corresponding ERU value using KDE, with a fallback on probabilistic bootstrapping.
If the ERU data of the bin has enough variation (distinct values), KDE is used to smooth the estimates for ERUs corresponding to new load values in that bin.
If the ERU data is sparse or has many duplicates, probabilistic bootstrapping is used by relying on frequencies of ERUs in that bin. Here, we resample ERU data points based on their probability of occurrence in that bin.

We group the data into $2000$ bins based on the node's RU levels such that each bin represents an \textit{equal percentile of total data}. Hence, for each bin, we store its node-level RU range and a distribution of the ERUs with their frequencies within that bin. 
To generate ERUs for a new policy based on node-level load, we identify the bin to which it belongs and use nonparametric estimation within that bin to generate corresponding ERUs using probabilistic sampling. This gives more weight to ERUs that occur more frequently for that bin.

\subsubsection{Empirical Tests of Convergence Guarantees}
\begin{figure}[t!]
    
    \centering
      \subfloat[Cluster 1a]{
        \includegraphics[width=0.45\columnwidth]{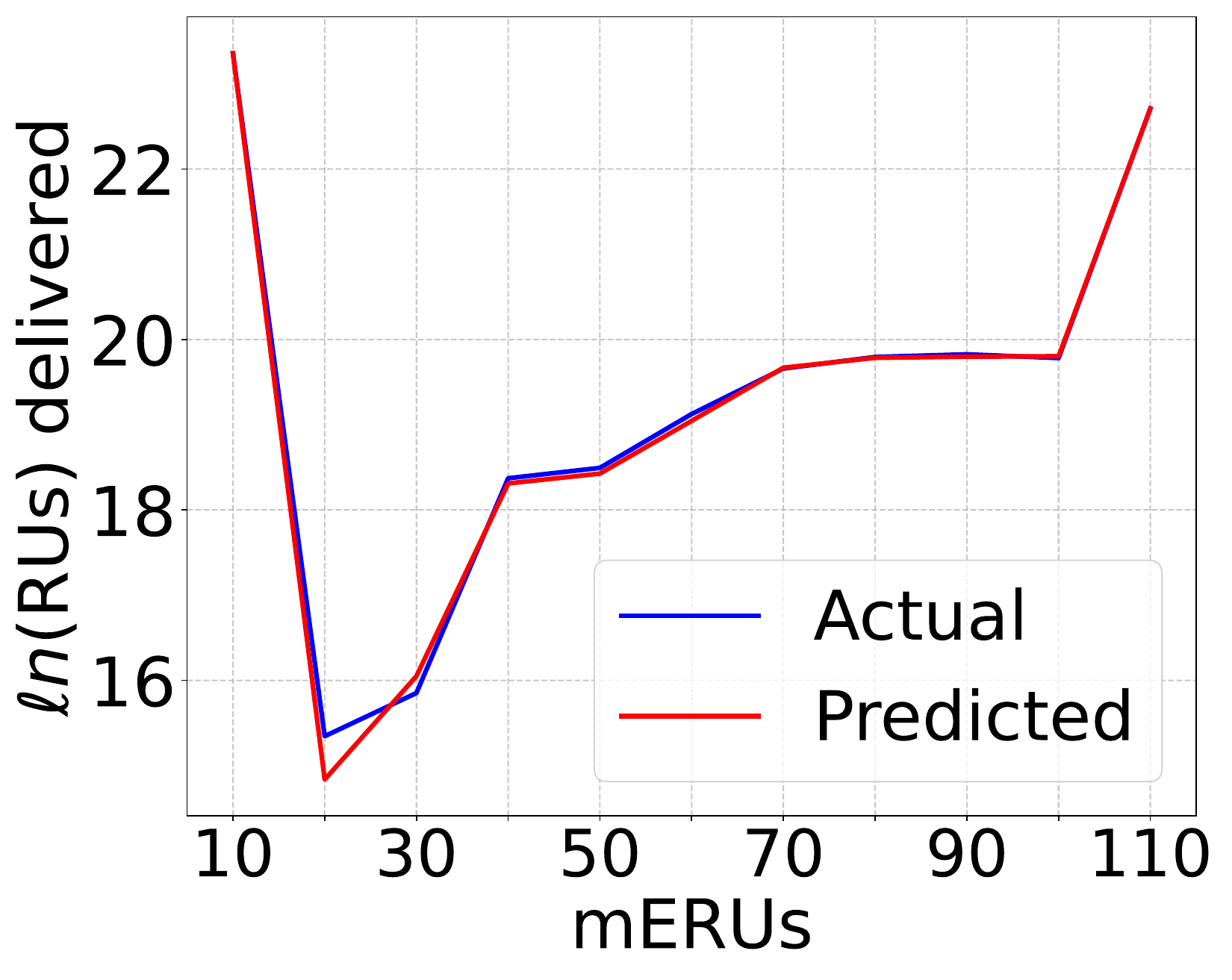}
        \label{fig:drvComparisonPredictedError_1a}
      }\quad
      \subfloat[Cluster 1b]{
        \includegraphics[width=0.45\columnwidth]{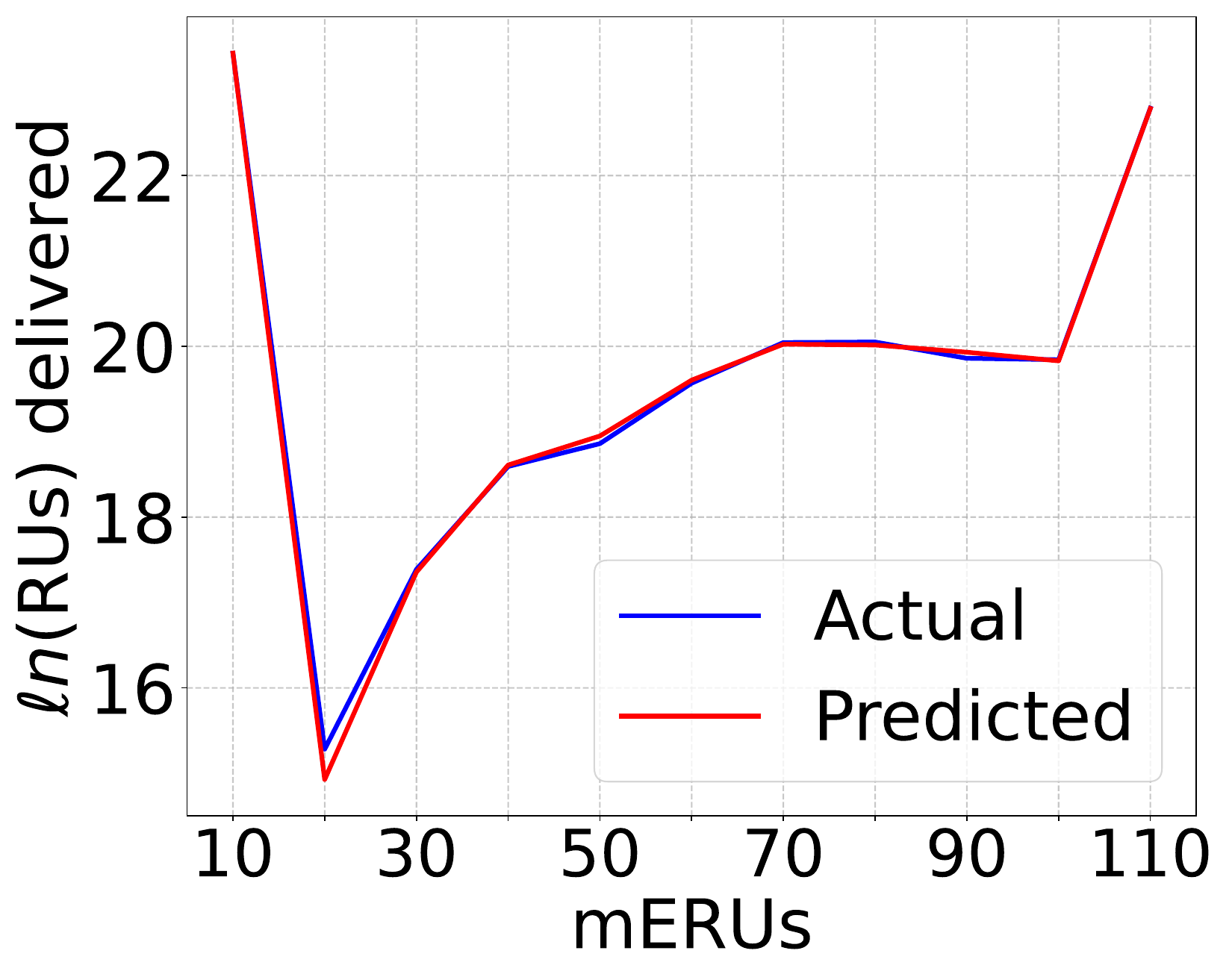}
        \label{fig:drvComparisonPredictedError_1b}
      }

    \caption{DRV profile of observed and predicted ERU values}
    \label{fig:drvComparisonPredictedError}
    
    \end{figure}

We compare the predicted ERUs with the observed ones for the same load levels, and show qualitative and quantitative distributional similarities. These are done on \textbf{C1a} and \textbf{C1b} datasets.
Figure~\ref{fig:drvComparisonPredictedError} plots the observed and predicted DRV profiles for 100k data points and their quantitative metrics are shown in Table~\ref{tbl:compStatMoments_combined}. The last column captures the \textit{$l_{1}$} norm of the prediction error. The \textit{$l_{2}$} norms of prediction error (RMS) are $0.1138$ and $0.1254$ for \textbf{C1a} and \textbf{C1b}, respectively. The substantial gap with \textit{$l_{1}$} error shows that the model is susceptible to outliers and the prediction errors in boundary regions (high ERUs) may be high. However, when used to validate system reliability, all ERUs beyond a certain level result in SLA failures and therefore their mutual distinctions are irrelevant. This is another reason why the last ERU bucket in DRV profiles is \textit{censured} with no upper bound.

\begin{table}[t!]
\centering
\small
\setlength{\tabcolsep}{3pt}
\renewcommand{\arraystretch}{1}
\caption{Actual and predicted ERUs for Clusters 1a and 1b.}
\begin{tabular}{L{1.25cm}||r|r||r|r||R{1.4cm}|R{1.4cm}}
\hline
\multirow[t]{2}{1.25cm}{\bf Statistical Measure} & \multicolumn{2}{C{1.5cm}||}{\bf Actual ERU
} & \multicolumn{2}{C{1.5cm}||}{\bf Pred. ERU
} & \multicolumn{2}{C{2.8cm}}{\bf $|$Observed - Predicted$|$}\\\cline{2-7}
 & \bf C1a & \bf C1b & \bf C1a & \bf C1b & \bf C1a & \bf C1b\\
\hline\hline
Mean & 0.0069 & 0.0086 & 0.0068 & 0.0087 & 0.0001 & 0.0001 \\
Median & 0 & 0 & 0 & 0 & 0 & 0 \\
Variance & 0.0050 & 0.0083 & 0.0044 & 0.0085 & 0.0006 & 0.0002 \\
\hline
\end{tabular}
\label{tbl:compStatMoments_combined}

\end{table}

Lastly, we apply this non-parametric model to example the effects of a placement policy. Specifically, Figure \ref{fig:drvComparisonPredictedError} plots the \textit{observed} DRV profiles in Cosmos DB for Cluster C1 using the Simulated Annealing policy, and the \textit{estimated} DRV profile for the same policy and cluster given by \simtool. 
The observed and estimated DRV profiles are almost indistinguishable. This is noteworthy since, in contrast, the scatter plot in Figure~\ref{fig:load-error-scatterplot} had poor correlations. But as claimed earlier, it was just an artifact of the granularity typical of statistical assessments; the ERU-RU model proposed in this section is able to learn those relationships.

\section{Forecast aware efficient packing policies with \policy}
\label{sec:packingpolicy}

In this section, we discuss the packing problem definition and the \policy PAM algorithm in detail. 
While we discuss the algorithm in reasonable detail, we keep the formal definitions concise and focus on the core decision points for the model.

\subsection{Load Balancing Problem Formulation}
\label{sec:loadBalancingProblem}

\subsubsection{Preliminaries}
Let $D^{i}$ indicate a database and $P^{i}$ be partitions of a database. $\mathcal{R}$ is the set of all \textit{partition replica instances} for partitions, with $R^i$ indicating a single replica instance. $\mathcal{M}$ is a cluster and $M^{i} \in \mathcal{M}$ are \textit{nodes} (VMs) in the cluster. $\mathcal{R}(M^i)$ is the set of replica instances hosted on a node $M^i$. In practice, all the nodes have identical resource capacities to ensure fungibility.

Let $I^{i} \in [0.0,1.0]^{d}$ be the normalized $d-$dimensional \textit{resource requirement vector} for a replica instance $R^i$. A value of $1.0$ indicates 100\% usage of that resource on a node. The key resource dimensions we consider are CPU, RAM, storage, read and write IO, and threads.
$I^{i}_{t}$ is the corresponding time-series object that captures the \textit{resource usage vectors}, until time $t$, for that replica instance. Each node has \textit{resource capacity constraints} on its total provisioning. So, the sum of resource requirement of replicas assigned to a node can never exceed $1.0$ for each of its dimensions, $\| \sum_{R^j \in \mathcal{R}(M^i)} I^j \|_{\infty} \leq 1.0$. In practical scenarios, cluster owners will put additional \textit{buffer} constraints on the node level provisioning, ensuring that the actual resource usage is even lower (e.g., CPU utilization is generally capped at 80\%). We will refer to this constraint vector as $\mathbf{B}$; the node level resource bounds can now be expressed as $\sum_{R^j \in \mathcal{R}(M^i)} I^j \leq \mathbf{B}$.

Let $C$ be the configuration matrix capturing the assignment of replicas to nodes; $\mathcal{C}^{i}_{j} = 1$ if replica $R^j \in \mathcal{R}$ is assigned to node $M^i \in \mathcal{M}$, and $0$ otherwise. Let $\mathcal{C}$ be the space of all these configurations. 
We define the \textit{PAM Policy} as $\Omega : \mathcal{C} \rightarrow \mathcal{C}$ that, given the current configuration $\mathcal{C}_{t}$ at time $t$ and the resource usage timeseries across all replicas $\mathcal{I}_{t}$ until now, finds the successor configuration $\mathcal{C}_{t+1}$ at time $t+1$ such that it has no capacity violations; for simplicity we assume such a feasible assignment always exists.
This includes three decisions: (1) Which replica to migrate, (2) When to move it, and (3) Which node to move it to. Migration of replicas between nodes and packing of replicas within a node are related: where and when to migrate replica workloads depends on the expected performance of the replicas (e.g., error rates for given workload) when packed on a node.
The policy will generate a new configuration only if the current configuration has violations, as discussed next. Treating $C$ as a set of (replica-node) assignments, the symmetric difference between the two configurations, $|C_{i} \Delta C_{j}|$ is the \textit{gap} between them, i.e., the 
replica migrations required to transition from $C_{i}$ to $C_{j}$.

Let 
$E^{i}_{t}$ be the total count of errors generated by node $M^i$ at time $t$. $E^{i}(I_{t})$ are the errors that will be generated by a node $M^i$ when it is assigned some load of $I_{t}$. These $I_{t}$ are $d-$dimensional resource vector which capture the \textit{total} aggregated resource requirements of all the replicas on a node. 
We have $^{C}E_{t}$ as the error generated jointly by all the nodes of the cluster in configuration $C$ at time $t$.

\subsubsection{Problem Definition}

Given a cluster $\mathcal{M}$, node level resource provisioning constraint $\mathbf{B}$, a set of replica instances $\mathcal{R}$ assigned to it, along with their $d-$dimensional resource usage time series, a starting configuration $C_{0}$, and a PAM policy $\Omega$. $\Omega(\mathcal{C},\mathcal{R})$ is a \textit{configuration path} in the space $\mathcal{C}$, the length of this path, denoted by $|\Omega(\mathcal{C},\mathcal{R})|$, is the number of times configuration change happened. Each of these changes may require multiple individual replica migrations, we refer to these as \textit{composite} migrations. 
The goal is to find the best policy $\Omega$ that optimizes the loss metric defined below among these configuration paths. 
Since each cluster is self-contained, this problem is independently solved for each.

We define a \textit{loss metric} $\mathcal{L}(\mathcal{M}, \mathcal{R}, \Omega)$ for the cluster specification $\mathcal{M}$, the instance loads for replicas $\mathcal{R}$, and the PAM policy $\Omega$ to capture the cost of executing a specific migration policy. 
Some candidate loss metrics for evaluating $\Omega$ are:
\begin{enumerate}[topsep=0pt]
    \item $|\Omega(\mathcal{R})| = $ the total number of \textit{composite migrations} 
    \item $ \sum_{0}^{|\Omega(\mathcal{R})|}|C_{i+1} \Delta C_{i}| = $ the total number of \textit{replica migrations} 
    \item $\sum_{t,j\in \mathcal{R}, C\in \Omega(\mathcal{R}), t\in T} {}^{C}E_{t}$ the \textit{total errors} emitted during the time period $T$. 
    \item  $\sum_{t,j\in \mathcal{R}, C\in \Omega(\mathcal{R}), t\in T} {}^{C}E_{t}.I^{j}_{t}$ the error level weighted sum of \textit{compute load delivered} during the time period $T$.
\end{enumerate}
Each of these metrics captures a key aspect of the load balancing costs. Optimizing the first one helps us contain the cost of actually computing the packing configurations repeatedly. In scenarios involving a large number of very small compute loads, calculating these configurations becomes expensive enough to preclude it from being done too frequently. 
The second metric, which minimizes the number of replicas migrated, mitigates the transient errors caused by migrations due to a temporary reduction in the RU serving capacity of the replica being moved. That said, migrations of replicas that are not currently serving high loads does not increase the error rates.
The last metric is a QoS type metric focusing on keeping the error rates as low as possible. DRV is a denormalized version of this function and plays a pivotal role in formalizing our \textit{reliable efficiency} objectives.

\begin{figure}[t!]

\small
\begin{algorithm}[H]
\caption{Triggering Re-balance \textit{(Phase 1)}}
\begin{algorithmic}[1]
\State \textbf{Input:} Configuration $C$, Predictions $P(I^{j}_{t+\Delta}) \forall j$
\State \textbf{Output:} $S$, List of nodes whose total load needs reduction
\For{every node $M^{i} \in \mathcal{M}$}
    \If{$\sum_{R^j \in \mathcal{R}(M^i)} I^j \nleq \mathbf{B}$} \Comment{\textit{Current Violation}}
        \State $S = S \cup \{M^{i}\}$    
    \EndIf
    \If{$\sum_{R^j \in \mathcal{R}(M^i)} P(I^j_{t+\Delta}) \nleq \mathbf{B}$} \Comment{\textit{Predicted Violation}}
        \State $S = S \cup \{M^{i}\}$    
    \EndIf
\EndFor
\State \textbf{return} $S$
\end{algorithmic}
\label{alg:trigger}
\end{algorithm}

\end{figure}

The \textit{policy optimization} problem is to find an $\Omega$ which minimizes the total loss $\mathcal{L}$ over a replica population $\mathcal{R}$.
\begin{equation}
    \Omega_{OPT} = \argmin_{\Omega} \mathcal{L}(\mathcal{M}, \mathcal{R}, \Omega)
\end{equation}
where, $\mathcal{M}$ is a \textit{fixed} cluster specification. We can say that the replica set $\mathcal{R}$ is sampled from an ideal distribution $\mathbb{R}$ to account for reasonable temporal variations in load time series. In our case, we have argued earlier that homogenizing utilizations leads to better QoS. In the following section, we propose a new model which achieves low P80/P20 ratios using load forecasts.

\subsection{Proposed Load Balancing Model}

Our \textsc{Optimizing Resource Balancing in Time (\policy)} load balancing module is composed of three decision phases, viz., triggering, selection and placement determination. \textit{Triggering} decides when to re-balance. These can be due to threshold breaches on resource capacity contentions, or more abstract reasons of attaining higher load uniformity across nodes. 
Once a trigger is provided, \textit{selection} determines which replicas should be migrated.
Finally, \textit{placement} decides where each replica should be placed. 
Various approaches for these three phases can be combined for an overall load balancing strategy. We next describe the design of our load balancing model which utilizes the quantile forecasts proposed in Section~\ref{sec:forecast}.

\subsubsection{Triggering Algorithm: When to Migrate Replicas out of a Node?}

We follow the standard node level constraint violation driven triggering policy with a simple generalization to accommodate the load forecasts. 
This algorithm marks a node for (out-)migration when either the current or predicted load (RU), CPU, memory, or storage utilization on a node violates capacity constraints. The capacity thresholds are usually set empirically based on past error behavior. Empirical proof of setting predictive windows is discussed in detail in the next section. 
This predictive mechanism aims to preemptively balance load, preventing a \textit{late} migration when the node and replica are already overloaded and serving traffic, which can cause errors to start occurring and continue on/become worse till the migration is completed.
Algorithm \ref{alg:trigger} formally defines these steps.

\begin{figure}[t]

\small
\begin{algorithm}[H]
\caption{Selection  of Migrating Replica \textit{(Phase 2)}}
\begin{algorithmic}[1]
\State \textbf{Input:} Unstable node set $S$, Predictions $P(I^{j}_{t+\Delta}) \forall j$
\State \textbf{Output:} $R_{S}$, List of replicas which should be moved out
\For{every node $M^{i} \in S$}
    \While {$\sum_{R^j \in \mathcal{R}(M^i)} I^j \nleq \mathbf{B}$ or $\sum_{R^j \in \mathcal{R}(M^i)} P(I^j_{t+\Delta})\nleq \mathbf{B}$}
        \If{$\sum_{R^j \in \mathcal{R}(M^i)} I^j \nleq \mathbf{B}$} \Comment{\textit{Current Violation}}
            \State $C = B_{k}$ \Comment{\textit{Pick resource dim. that is violated}}
        \EndIf
        \If{$\sum_{R^j \in \mathcal{R}(M^i)} P(I^j_{t+\Delta}) \nleq \mathbf{B}$} \Comment{Future Violation}
            \State $C = B_{k}$ \Comment{\textit{Pick resource dim. that is violated}}
        \EndIf

\State $R^{j} = \Call{BinaryFirst}{M^{i}, C}$ \Comment{\textit{Pick smallest replica that brings us within threshold}}
        \State $R_{S} = R_{S} \cup \{R^{j}\}$    
        \State $\mathcal{R}(M^i) = \mathcal{R}(M^i) \setminus R^{j}$
    \EndWhile
\EndFor
\State \textbf{return} $R_{S}$
\end{algorithmic}
\label{alg:selection}
\end{algorithm}

\end{figure}

\subsubsection{Selection Algorithm: Which Replica(s) Should Be Migrated from a Node?} 
The various selection policies vary from lowest resource usage (minimize end user disruption) to the highest (minimize migration counts)~\cite{farahnakian2016energy}.
To balance the user disruption and migration cost goals simultaneously, we instead take an intermediate approach.
For every node marked for load reduction by the triggering phase, we arbitrarily choose a member from the set of violated resource constraints and execute a binary search on an ordered (by the same resource) list of replicas to identify the \textit{smallest replica set} that brings the breach below the threshold. We refer to this subroutine as \textbf{BinaryFirst}, with details given in Algorithm~\ref{alg:selection}.

\begin{figure}[t]

\small
\begin{algorithm}[H]
\caption{Placement of Migrating Replica \textit{(Phase 3)}}
\begin{algorithmic}[1]
\State \textbf{Input:} Unstable node set $S$, Migrating replica set $R_{S}$
\State \textbf{Output:} $\Omega:R_{S} \rightarrow \mathcal{M}$, map of  replicas migrating into cluster
\For{replica $R^{j} \in R_{S}$, in decreasing order of median RU size}
    \State $M = \Call{PolicyFit}{R^{j},~P(I^{j}_{t+\Delta})}$
    \If{$M == NULL$}
        \State $M = \Call{PolicyFit}{R^{j}}$
    \EndIf
    \State $\Omega(R^{j}) = M$
    \State $\mathcal{R}(M) = \mathcal{R}(M) \cup R^{j}$  \Comment{\textit{$R^{j}$'s load added into $M$ for future decisions}}
\EndFor
\State \Return{$\Omega$}
\end{algorithmic}
\label{alg:placement}
\hrule
\begin{algorithmic}[1]
\Procedure{PolicyFit}{$R,~  P(I_{t+\Delta})$} 
\State \textbf{Input:} A replica $R$ and its load forecast $P(I_{t+\Delta})$
\State \textbf{Output:} Node $M$ with the best score
\For{node $M^{i} \in \mathcal{M}$ which can accommodate $R$}
    \State fit = \Call{WorstFitScore}{$M^{i}$} \Comment{\textit{Empty load capacity available normalized by the maximum empty space}}
    \State $\text{skew} = -\frac{1}{\pi}\arccos \frac{\langle I(P(M^{i} \cup I_{t+\Delta})), \mathbb{1} \rangle}{|I(M^{i} \cup P(I_{t+\Delta}))||\mathbb{1}|} $
    \State $\text{score[i]} = \text{fit} + \alpha.\text{skew}$ 
    \If{$score[i] > max\_score$}
        \State $max\_score = \text{score[i]}$
        \State $max\_idx = i$
    \EndIf
\EndFor
\State \Return{$M^{max\_idx}$}
\EndProcedure	
\end{algorithmic}
\label{alg:placementpolicy}
\end{algorithm}

\end{figure}

\subsubsection{Placement Algorithm: Where to migrate the replica(s)?}
The third and last phase of load balancing determines the optimal destination node for every replica being moved. 
Our proposed algorithm uses a modified \textit{worst-fit approach} that incorporates two scoring metrics that are combined linearly for a final score.
\begin{enumerate}[topsep=0pt,leftmargin=*,itemindent=15pt]
\item\textit{Load Factor:} The normalized available space on the node, calculated as the node's unused RU capacity divided by the maximum empty space across all nodes. This score captures the 'worst-fit' aspect of our policy and biases the selection towards \textit{emptier} nodes.

\item \textit{Skewness:} This is a key component in all vector packing strategies~\cite{panigrahy2011heuristics,christensen2017approximation,eigen23} that helps reduce the original high-dimensional vector packing problem to a lower dimensional tractable version, while possibly losing some accuracy. We define skew as,
\begin{equation}
\text{skew} = -\frac{1}{\pi}\arccos \frac{\langle I(M^{i}), \mathbf{1} \rangle}{|I(M^{i})||\mathbf{1}|}
\end{equation}
Skew measures the angle between a node's normalized resource utilization vector $I(M^{i})$ and the symmetric utilization vector (with all 1's).
The minus sign and scaling normalize the value in the right direction. The policy goal is to find placements that minimize the angle. This helps avoid creation of node configurations with very skewed resource availabilities. E.g., a node with no storage available is unlikely to able to fill in its remaining CPU/RAM capacity as every replica will need some basic storage footprint.
\end{enumerate}

The placement algorithm linearly combines these scores to generate a final worst-fit ranking for all potential destination nodes. The scale factor $\alpha$ used for combining the two scores is determined empirically in offline tests. Refer to Algorithm \ref{alg:placement} for a formal outline of the \policy placement algorithm.
These policies depend on a suitable time series load forecasting model, which we propose next.

\section{Load Forecasting with \forecast}
\label{sec:forecast}

Mapping workloads to the available resources is a crucial scheduling problem to meet the required QoS.
This reduces to an online stochastic bin packing problem which, in its most general form, is computationally intractable~\cite{panigrahy2011heuristics}. However, these are worst-case adversarial bounds and in cases like managed database services with long customer histories of compute requirements, \textit{load forecasting} can potentially help overcome the complexity barriers and achieve better approximation ratios, as seen in earlier~\cite{eigen23,farahnakian2016energy}. Both the data context and the packing algorithm which will use the forecast are equally important in determining the outcome.

Specifically, the goal of our forecasting model is to \textit{predict the demand for resources,} CPU, RAM, Storage and load (RUs). Formally, the \textit{target variable} for the prediction is the resource usage $P(I^i_{t+\Delta}|I^i_{t-n})$ for a given temporal look-ahead window $\Delta$, given the historical usage of $n$ time units.
Several salient aspects of our context govern our forecasting framework's design are:
\begin{itemize}[leftmargin=*]
    \item \textsc{Low Granularity Forecasts:} 
    Since every node has substantial buffer capacity to support transient RU bursts, 
    we ignore all volatility at time-granularity finer than 1-minute, and capture only systematic variations happening over 10s of minutes.

\item \textsc{Reasonable Forecast Window:} Frequent replica migrations impacts both reliability and efficiency.
    A reliable forecast of, say, 2 hours into the future can help avoid consecutive migrations within that time window.

\item \textsc{Asymmetric Losses}: Underestimating RU requirements impacts reliability due to over-packing of nodes while overestimating it causes efficiency loss due to under-used capacity. The optimal tradeoff point between reliability and efficiency is a product decision. 
    The ideal forecasting model should work well on a wide range of these choices.

\item \textsc{Point vs. Distribution Estimates:} While most predictive load balancing works rely on using good point forecasts~\cite{eigen-vldb-2023, farahnakian2016energy}, they can be costly to compute. Instead, we posit that distributional forecasts are \textit{much more robust} and provide sufficient information for a compatible packing policy.

\item \textsc{Spatial Scope:} Our load forecasts are at the granularity of each replica instance, $i$, as required for the load balancing formulation in Section~\ref{sec:loadBalancingProblem}.
\end{itemize}

\textsc{Load Utilization prediction for NoSQL workload Analytics (\forecast)} is simple predictive model that incorporates the above principles. The time series modeling problem is simplified to predicting the \textit{peak load value} over 30~minute blocks. We use asymmetric loss functions that have a higher over-estimation penalty than under-estimation, implying reliability is more important than efficiency. Finally, we train one model for one specific quantile to keep the model complexity low.

We implement a  time series forecasting framework for replica loads using Microsoft's \textit{LightGBM} with quantile regression. We use \textit{Whittaker-Eilers smoothing} to smooth replica-level RU time series data, helping reduce noise while maintaining the trends in the data. 
The timestamps are grouped into \textit{30-minute intervals}, ensuring a consistent time granularity for the prediction task.
\textit{Feature engineering} is driven by basic statistics, and includes time-based features such as the day of the week, hour of the day, and minute of the hour, rolling window statistics (mean, median, and quantiles) over multiple time horizons, along with higher-order moments (skewness and kurtosis) that help capture the temporal structure and volatility in the time series data.
LightGBM's quantile regression uses Pinball Loss to predict specific percentiles of the time series.

\subsection{Evaluation of \forecast}
Empirical tests are performed on the \textbf{C1b} workload, with 6 days of training data and another 6 days as test data. We first analyze the performance of predicting the total load (RUs) and benchmark against standard forecasting models. We then introduce a simple \textit{coverage} metric relevant to our context that helps better visualize the performance of \forecast's forecasts. We then briefly discuss the coverage behavior of RU and resource predictions.

\subsubsection{Overall Load (RU) Prediction}
We first evaluate \forecast for RU predictions. We measure the Pinball Loss (quantile loss)~\cite{steinwart2011estimating} of these predictions to assess their statistical accuracy and compare our model with two classical time-series forecasting techniques:\textit{ARIMA}, which trains a stationary time series model with seasonal adjustment enabled, and with autoARIMA for parameter tuning;
and \textit{Exponential Smoothing (ETS)} trained with a smoothing parameter of 0.2.
Table~\ref{tab:load_comparison} shows that our model has a much better accuracy than both ARIMA and ETS across load percentiles.

\begin{table}[t]
  \centering
  \small
  \caption{Comparison of Our Forecasting Model with ARIMA and ETS (Pinball Loss) on RU Load for C1b}
  
  \label{tab:load_comparison}
  
  \begin{tabular}{c|r|r|r||rr}
    \toprule
    &  &  &  & \multicolumn{2}{c}{\bf Improvement Over} \\
    \textbf{Percentile} & {\textbf{\forecast}}
& \textbf{ARIMA} & \textbf{ETS} & \textbf{ARIMA} & \textbf{ETS} \\
    \midrule
    0.001 & 58 & 3069 & 3751 & 98.1\% & 98.5\% \\
    0.25  & 1080 & 2800 & 3360 & 61.4\% & 67.9\% \\
    0.5   & 1420 & 2531 & 2968 & 43.9\% & 52.2\% \\
    0.75  & 1323 & 2262 & 2576 & 41.5\% & 48.6\% \\
    0.9   & 953  & 2100 & 2340 & 54.6\% & 59.3\% \\
    0.995 & 349   & 1997 & 2191 & 82.5\% & 84.1\% \\
    \bottomrule
  \end{tabular}

\end{table}

To better understand \forecast's performance, we introduce a simple visual aid metric. \textsc{Coverage Deviation} $C(p_{1}, p_{2})$ is the gap between the expected and the observed proportion of times when the actual load (RUs) $l_{t}$ is falls between those percentile values, i.e., $C(p_{1}, p_{2}) = |\{t : l_{t} \in [L(p_{1}), L(p_{2})]\}| - |p_{2} - p_{1}|$. Here $t$ is a \textit{normalized} time between $0$ to $1$ and $L(p_{i})$ is the predicted $i^{th}$ percentile of the load (RU) value. The sign of this metric indicates in which percentile ranges the forecasts are systematically under/over-estimating. 
Figure~\ref{fig:quantile_errors} shows the coverage deviations for the RU predictions by \forecast. The highest values appear in the top left region, which means that most of data appears between the lower percentile bands, e.g., $[.01,.25]$. This shows that our model is systematically \textit{overestimating the loads}. An equivalent observation follows from focusing on the lowest negative value in the table and follows the same reasoning.

\subsubsection{CPU/RAM/Storage Prediction} 
\begin{figure}[t]

    \centering
    \includegraphics[width=\columnwidth]{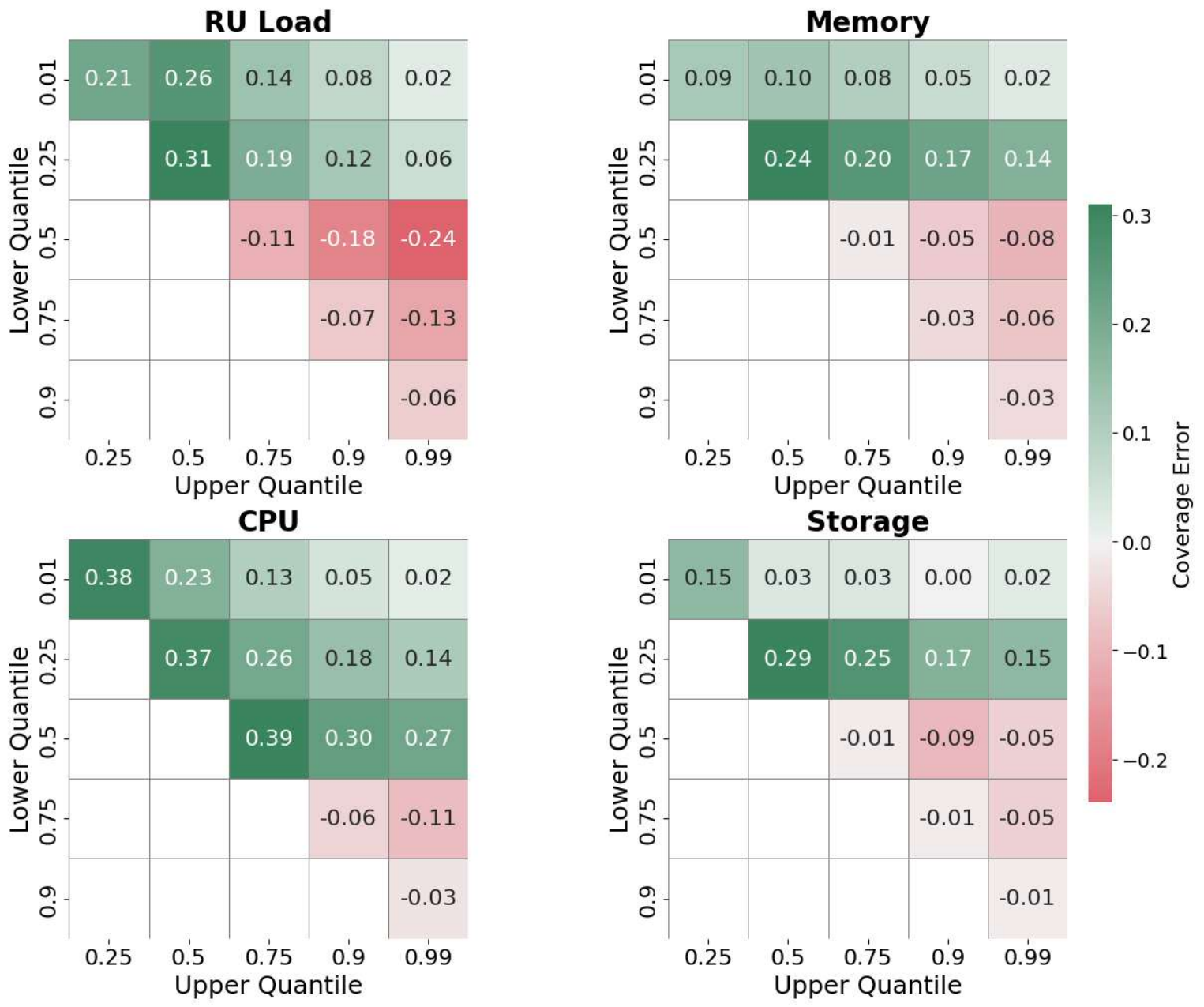}
    
    \caption{Coverage Deviations for RU Load, Memory, CPU, and Storage. Positive values indicate over-coverage; negative values indicate under-coverage for C1b.}
    \label{fig:quantile_errors}

\end{figure}

We repeat the same analysis with resource-level time series predictions. Figure~\ref{fig:quantile_errors} illustrate the coverage deviations of \forecast on RAM, CPU, and Storage. While the pattern is similar to RU, that values are less extreme. So the model has an overestimation bias for all the resources' cases but it is weaker than that of RU. 
Table~\ref{tab:load_comparison} shows that the absolute performance numbers, by themselves, it is not enough and the true test 
comes from it effectiveness for actual packing scenarios.

\section{Results on PAM Policy}
\label{sec:results}

We evaluate our proposed \policy PAM policy based on the \forecast prediction models
using our \simtool policy simulator, for the four different Cosmos DB cluster datasets we have provided.
We first compare our \policy policy against the existing Simulated Annealing (SA) method used in Cosmos DB and a baseline policy based on a Worst-Fit Optimization (WFO), primarily using 6 days of \textbf{C1b} workload for a detailed analysis (Sec.~\ref{sec:results:compare}).
We then show the generalizability of
our approach to the other clusters, \textbf{C1a, C2, C3} (Sec.~\ref{sec:results:generalize}). Next, we examine the reduction in nodes possible using \policy without significantly impacting the error rates (Sec.~\ref{sec:results:nodes}).
Lastly, we discuss hyper-parameter tuning of the models (Sections~\ref{sec:results:skew},~\ref{sec:results:foresight} and~\ref{sec:results:forecast}).

\subsection{Comparing with Default \& Baseline Policies}
\label{sec:results:compare}

Besides the default SA algorithm used by Cosmos DB (Sec.~\ref{sec:sa}),
we design a non-trivial no-forecast baseline using a Worst-Fit Online (WFO) packing policy to illustrate the hardness of this problem. 
WFO uses constraint programming to formulate replica placement as a multi-dimensional bin-packing optimization problem~\cite{V_Romero_2023} across the 4 resource dimensions of RU load, CPU, memory and storage.
In each time period,
it operates on each replica sequentially based on its current load, and chooses a node that it should be placed on based on maximum available capacity, i.e., worst fit.
It preserves existing replica assignment to nodes across consecutive time periods if no change is required,
and optimizes bin-packing only for new replicas, with constraints updated by the removal of departing replicas. 
When incremental packing fails due to resource constraints, it triggers an expensive full-repack.
We implement WFO using Google's OR tools~\cite{ortools}.
For \policy, we use default hyper-parameters of $\alpha=4$, $2h$ ahead forecast window and forecast percentile of $P50$ median from the \forecast model's output.

\begin{figure}[t!]
    \centering    \includegraphics[width=\columnwidth]{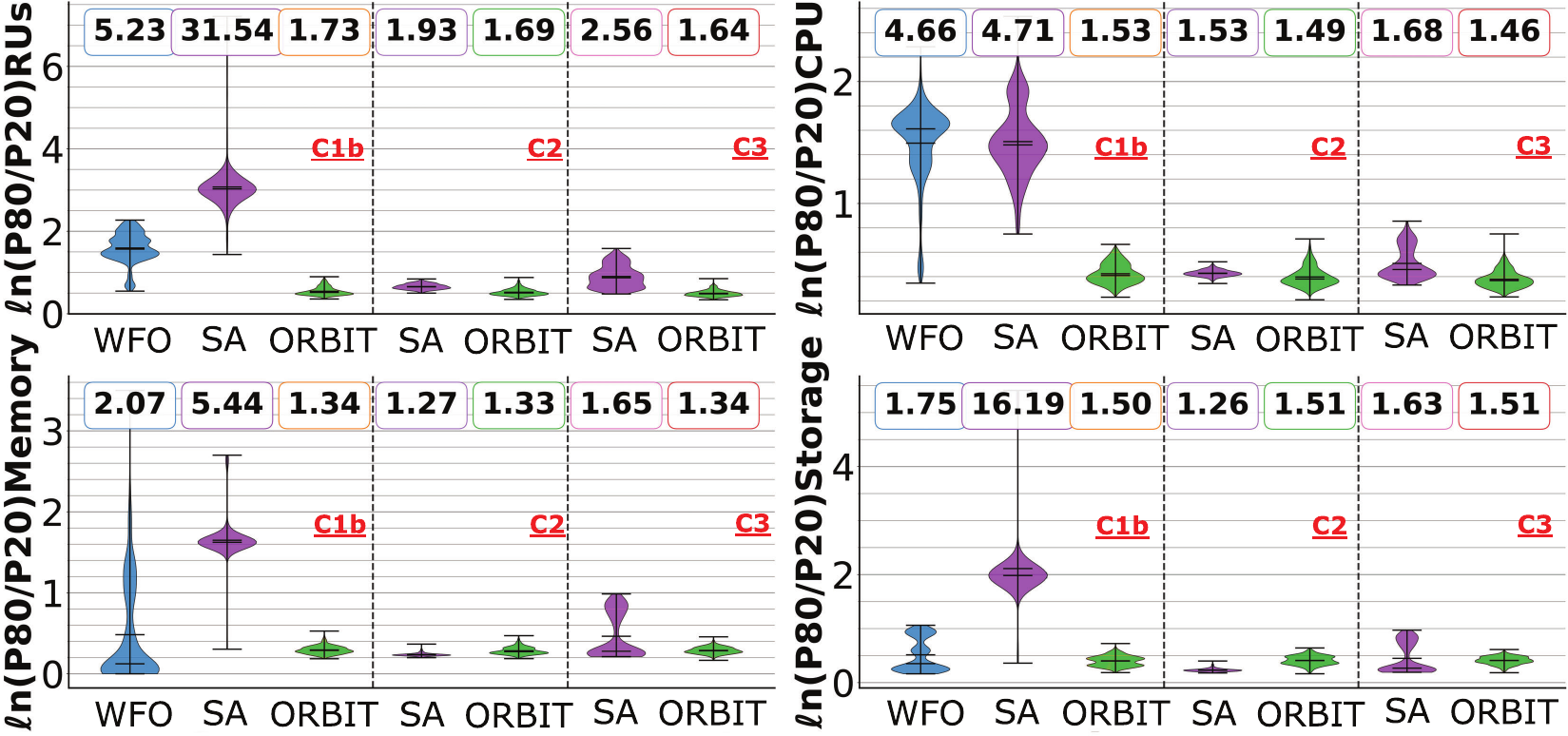}
    
    \caption{Comparison of $\ln\Big(\frac{P80}{P20}\Big)$ ratios between \policy, WFO and SA.
    Mean of load (non-ln) is shown in label.}
    \label{fig:log_comparison_c1c2c3_violin:2}
    
\end{figure}

Fig.~\ref{fig:log_comparison_c1c2c3_violin:2} (left) shows violin plot distributions of the natural log of the P80/P20 ratios for the four resource metrics for Cluster C1b, sampled every 5~mins, when this workload is packed using WFO (blue), SA (purple) and \policy (orange), as reported by the \simtool policy simulator. As we see, the ratios for WFO are better than SA, with lower variability and also resource ratios that avoid hotspots in the cluster. This confirms that WFO is a competitive baseline, and yet \policy does even better than WFO with an even tighter and lower violin, with the median P80/P20 ratio ranging from 1.34--1.73 across the resource types. This shows substantial \textit{improvement in the uniformity of utilization across nodes}, with little difference between cold (P20) and hot (P80) nodes. 
This improvement carries forth to C2 and C3 as well, with \textsc{Orbit} having lower ratios than SA; WFO could not complete on these larger clusters due to high time complexity.
Using alternate ratios such as P90/P10 and P70/P30 give similar rank ordering for these policies.

\begin{figure}[t!]

    \centering
      \subfloat[Comparison across strategies.]{
        \includegraphics[width=0.48\columnwidth]{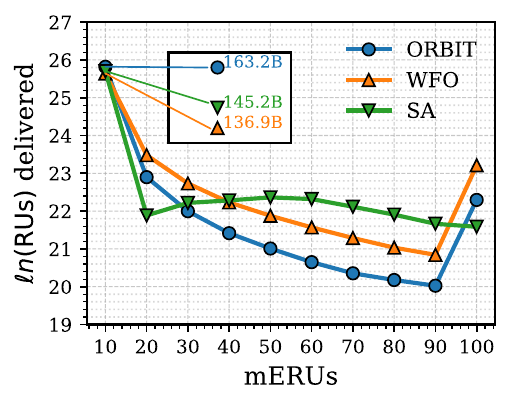}
        \label{fig:drv-sa-baseline-comparison}
      }
      \subfloat[Node footprint \% provided, with \policy.]{
        ~~~\includegraphics[width=0.48\columnwidth]{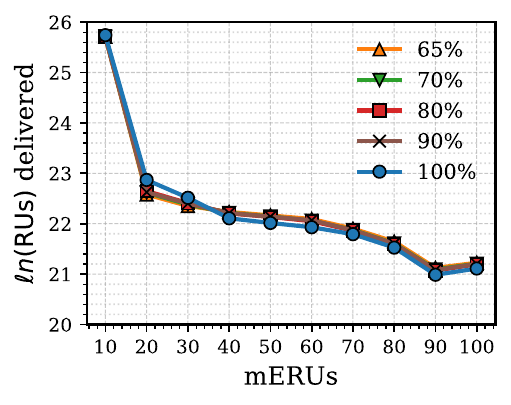}~~~
      }
      
    \caption{DRV profile for different strategies and hyper-parameters on Cluster C1b}
    \label{fig:drv-footprint}
    
\end{figure}

This consequently improves the DRV profile of placements done using \policy (Fig.~\ref{fig:drv-sa-baseline-comparison}), which exhibits higher RUs delivered at lower ERUs, compared to SA and WFO. E.g., for the lowest error of 10~mERUs (Fig.~\ref{fig:drv-sa-baseline-comparison}, inset), \policy delivers 163B RUs, compared to 137B for WFO and 145B for SA. Since the total load is the same for all policies, the reliability is highest for \policy.

The total number of migrations over 6 days is 19.3k for WFO, 34k for SA and just 47 for \policy; but in the context of 35.4k total replicas, these are all modest migration numbers. While WFO accommodates the replica flux incrementally without requiring any regular migration, it does causes a full re-balance once during this period with 19.3k migrations taking place along with the associated flux; this peak is 18.7k migrations for SA and 7 for \policy.

WFO cannot be used for online decisions due to its latency to solution. E.g., when $<500$ replicas arrive/depart in a 5-min window, it takes 30--60s to solve, but this rises to 5--8~mins for $3000+$ replicas and almost 1h for a full repacking. In contrast, \policy takes under 30s to execute 
but, importantly, does not need to execute in real-time since it relies of \forecast's forecasts that give it 10s of minutes if not hours of lead time to decide and execute the re-balancing.

\subsection{Generalizability to Diverse Clusters}
\label{sec:results:generalize}

Next, we demonstrate the generalizability of the \policy to different clusters, validating it 
over the bigger and qualitatively different cluster datasets, \textbf{C2} and \textbf{C3}. 
In Fig.~\ref{fig:log_comparison_c1c2c3_violin:2}~(right violins), we see that the P80/P20 ratios for our policy remains low using \policy for C2 and C3 (green and red violins), showing a comparable if not better median ratio than even C1b with resource ratios ranging between 1.33--1.69 and 1.34--1.64, respectively. This confirms that our combine forecasting and policy approaches (\forecast+\policy) together are able to achieve a more uniform resource usage across all nodes, even with a longer time period of evaluation (54 days), and avoid hotspots that can lead to user-facing long-tail errors.
The hyper-parameters for \policy were themselves chosen based on a ablation study for the Cluster C1b, as discussed later, and tuning on such smaller datasets is still sufficient to generalize to larger clusters. The baseline WFO policy is not reported for these clusters due to the high time complexity (running into days/weeks) to solve for the large time periods for these datasets.

\subsection{Resource Efficiency Gains}
\label{sec:results:nodes}

\begin{figure}[t!]

    \centering    \includegraphics[width=\columnwidth]{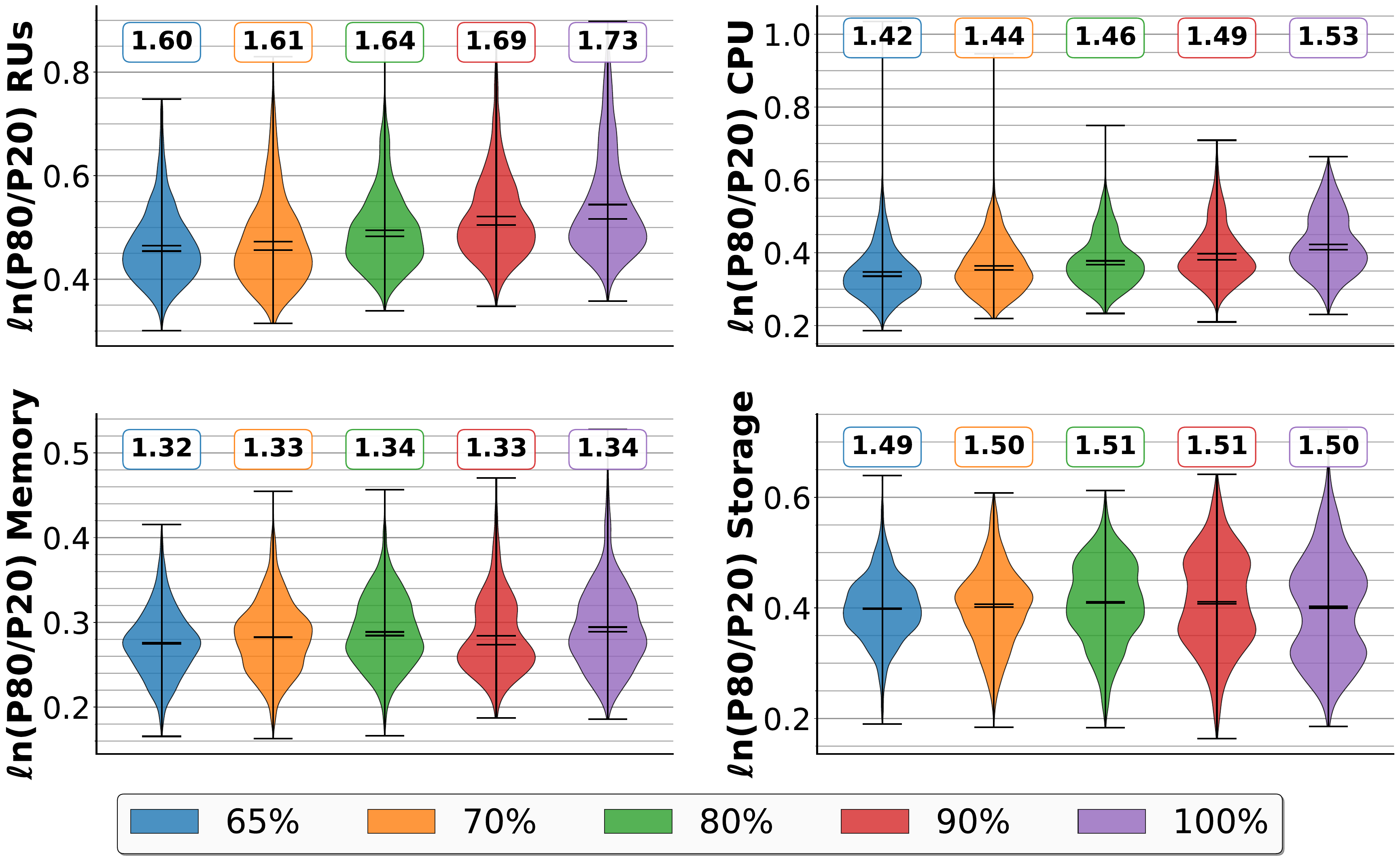}
    
    \caption{$\ln\Big(\frac{P80}{P20}\Big)$ ratio for each resource for nodes packed using \policy for different \textit{node footprints}, for Cluster C1b}
    \label{fig:logp80p20_nodefootprint}
    
\end{figure}

While the earlier results showed the improvement in tail errors and better user experience using \policy, we use the same framework to explore potential reduction in resource allocation for then clusters to reduce the operational costs. 
Specifically, we \textit{reduce the number of nodes} assigned to the C1b cluster in steps of 5\%, from 100\% to 65\%, and use our \policy for the placements; any lower and the replicas could not be placed. We report their DRVs in Fig.~\ref{fig:drv-footprint} and P80/P20 ratios in Fig.~\ref{fig:logp80p20_nodefootprint}.
As we see, as the resource footprint reduces, the P80/P20 resource deviations marginally change, e.g., from 1.73 for RU ratios with 100\% nodes to to 1.60 for 65\% nodes, which is actually an improvement.
We also see that the DRV profiles are comparable even with fewer resources, with some higher RUs at higher error rates,
e.g., delivering 151.2B RUs using all 100\% nodes vs. a lower 147.5B RUs with only 65\% nodes, at a low error rate of 10~mERU, and 1.47B RUs with 100\% nodes vs. 1.65B RUs with 65\% nodes, at a higher error of 100~mERU.
Migrations caused remain small enough
to not play a factor.
Since the node resources are captively allocated to the cluster, a reduction in the node footprint directly results in \textit{lower operational costs} for Cosmos DB.

\subsection{Tuning Skewness Parameter $\alpha$}
\label{sec:results:skew}

Next, we outline the effect of different hyperparameter tuning on \policy, with the best ones chosen in the results reported above.

We first compare different values of $\alpha$, which is the weightage given to the \textit{skewness} in Algorithm~\ref{alg:placementpolicy}.
This is to assess the relative impact on PAM quality measured in terms of total migrations, utilization ratios, and skew distributions. Here we use a ``perfect'' (oracle) forecast since the goal is to select the most suitable value of $\alpha$ without biases introduced by our \forecast forecasting model. Hence, we evalaute these on the first 6 days of the C1b dataset.

\begin{figure}[t]

    \centering \includegraphics[width=\columnwidth]{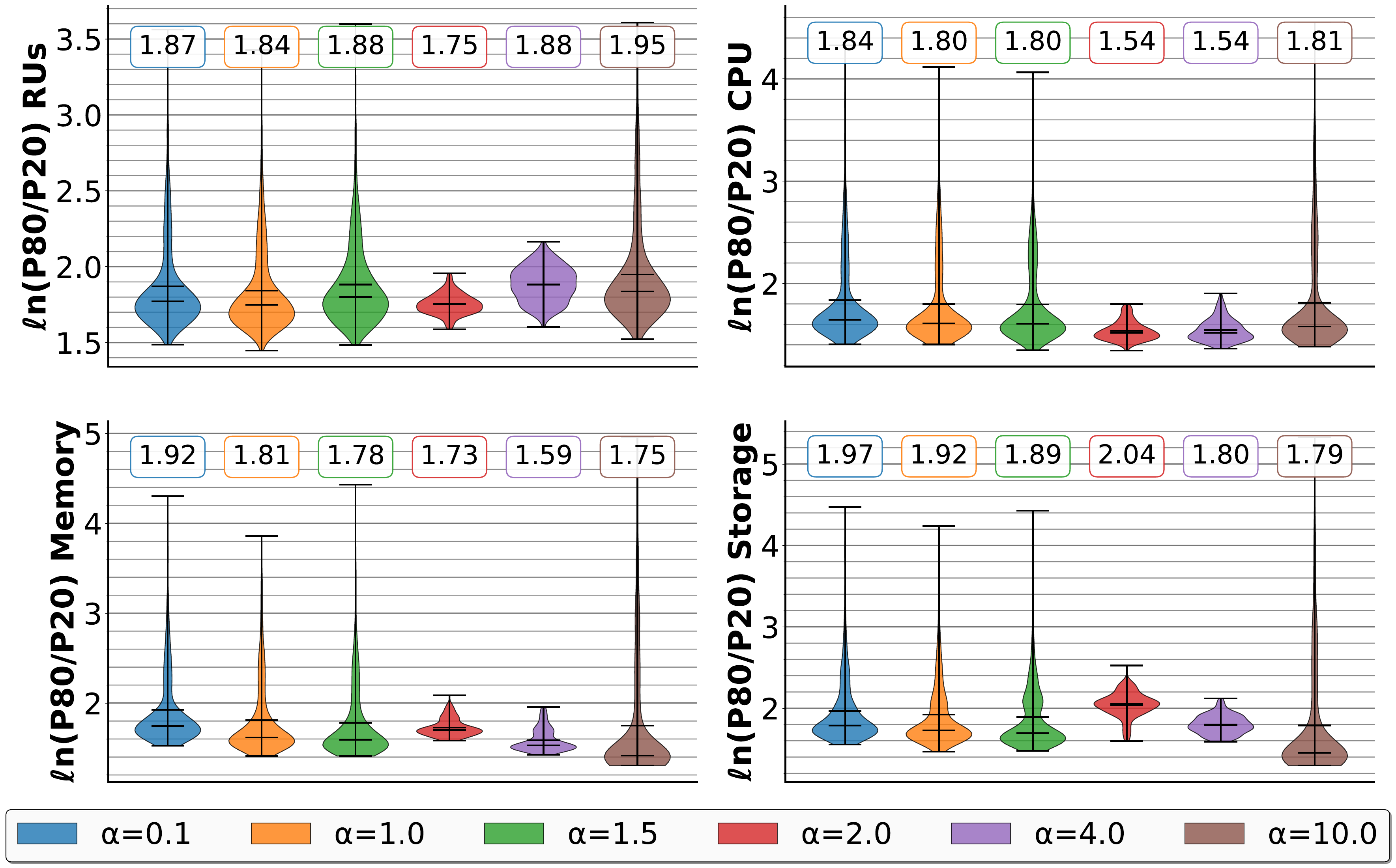}
    
    \caption{$\ln\Big(\frac{P80}{P20}\Big)$ ratio for each resource for nodes packed using \policy at different \textit{$\alpha$ skewness}, for Cluster C1b}
    \label{fig:tAlphaTuning}
    
\end{figure}

We try values of $\alpha$ ranging from $0.1$ to $10.0$, and observe their P80/P20 violins (Fig.~\ref{fig:tAlphaTuning}). Values of skewness that cause upper long-tails and outliers are not suitable since they indicate hotspots among the nodes.
As we eliminate such $\alpha$ values we are left with values in the middle ranges of Fig.~\ref{fig:tAlphaTuning}. 
The $\alpha$ skewness also modestly affects the number of migrations that are triggered by our \policy policy,
but the counts are small enough ($<10$) in all cases to not materially affect the performance.
The DRV profiles at different $\alpha$ (not shown) are similar between 1.0 and 4.0, but worse at the extremes of 0.1 and 10.
Hence, we 
fix the default as $\alpha=4$ in all experiments.

\subsection{Tuning Forecast Window for PAM Policy}
\label{sec:results:foresight}

We next evaluate various forecast time window sizes, i.e., how long ahead we forecast the resource loads using \forecast model before passing it to the \policy model. Typically, longer forecasts can lead to more stable packing but may also have higher forecasting errors.
Forecasting noise can translate to more resource contention.

\begin{figure}[t]
    \centering \includegraphics[width=\columnwidth]{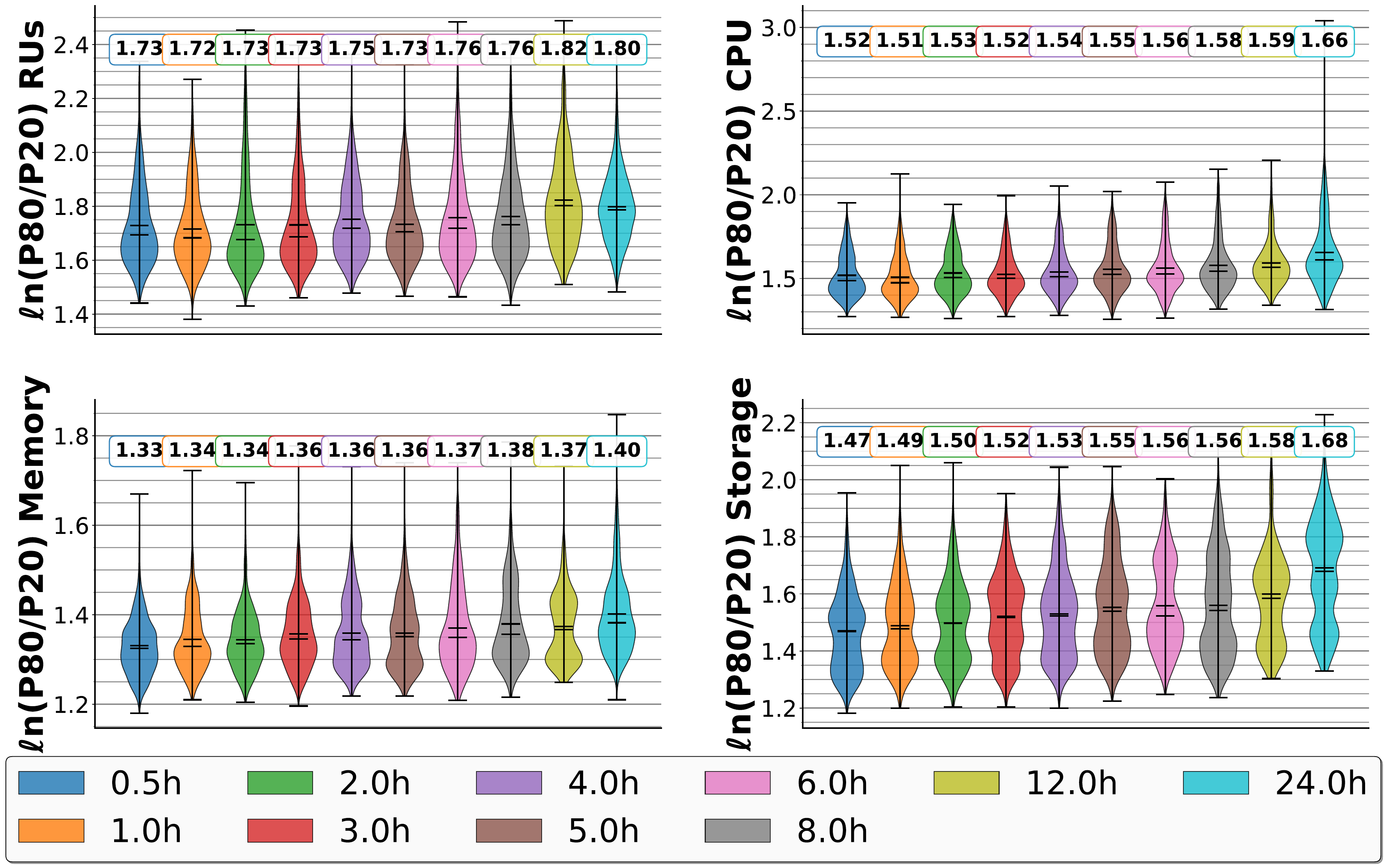}    
    \caption{$\ln\Big(\frac{P80}{P20}\Big)$ ratio for each resource for nodes packed using \policy at different \textit{forecast windows}, for Cluster C1b}
    \label{fig:forecastWindows}
\end{figure}

We evaluate time windows of 0.5h to 24h forward time windows, and include the \textit{zero} time window as well, to be able to compare with the vanilla no-forecast version of the policy as a baseline.
Fig.~\ref{fig:forecastWindows} reports the $\ln$ of the P80/P20 ratio for the resources.
The first result that stands out is that not using forecasts (corresponding the 0 hr time window) creates multiple orders of magnitude of degradation in first few days and it continues to be poor even later.
Secondly, among the different forecast windows $>0h$, the relative differences in P80/P20 ratios and even the DRV profiles (not shown) are small, which indicates that even a 30 minutes advance knowledge of upcoming load can play a key role in better utilization ratios.
Lastly, the performance of these models are not sensitive to small parameter changes, which makes them more generalizabile.

As a result, based on the modestly lower P80/P20 ratios for RUs and fewer outliers for other resources, and with an intent to lower the cost and potential errors when computing longer windows, the $2h$ forecast window is used as the default parameter for \policy.

\subsection{Tuning Forecast Percentile}
\label{sec:results:forecast}

\begin{figure}[t]
    \centering \includegraphics[width=\columnwidth]{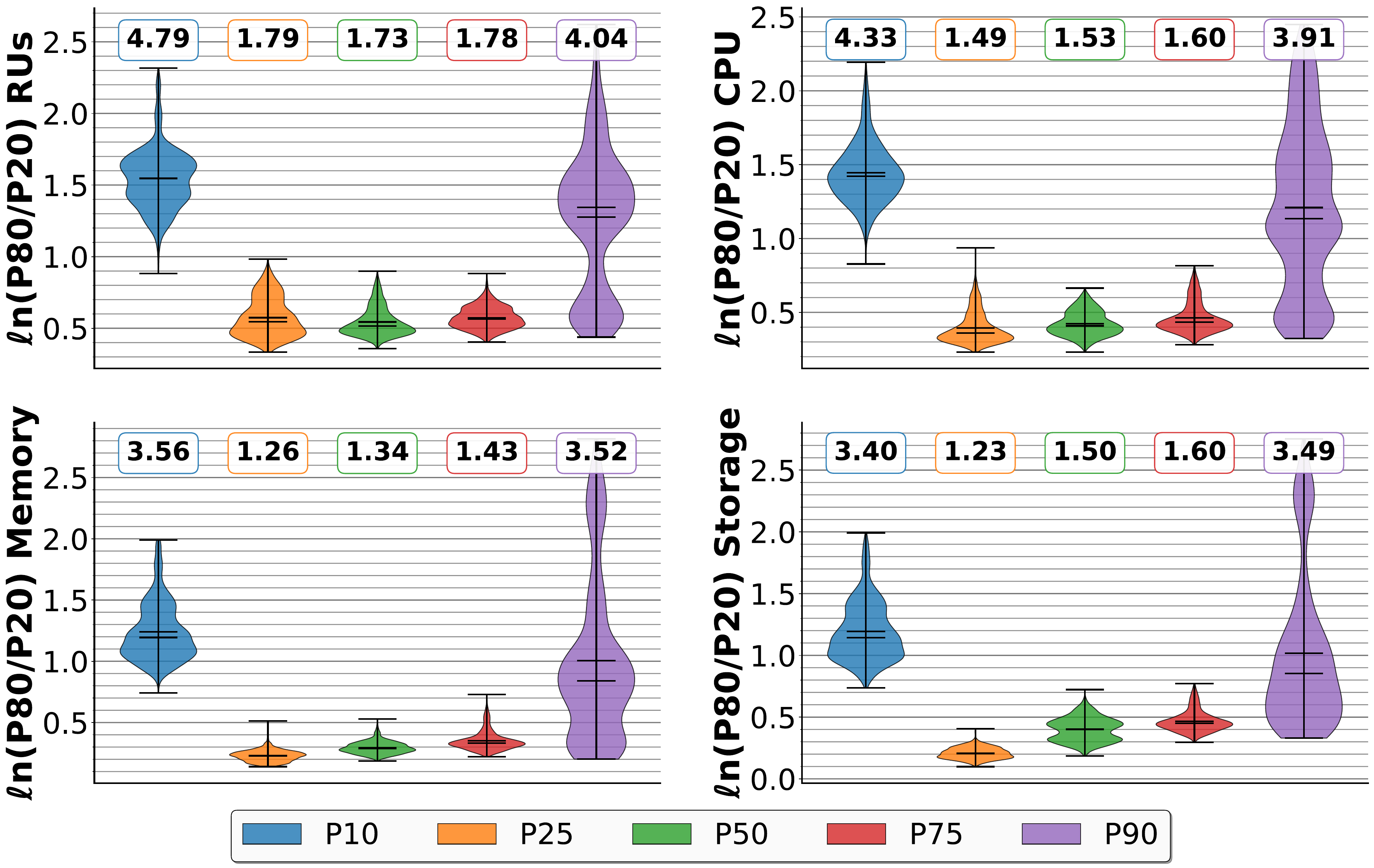}
    
    \caption{$\ln\Big(\frac{P80}{P20}\Big)$ ratio for each resource for nodes packed using \policy at different \textit{forecast percentiles}, for Cluster C1b}
    \label{fig:logOfConservativenessViolin}
    
\end{figure}

As seen in Section~\ref{sec:forecast}, the \forecast forecasting model is tuned conservatively to have a higher under-estimation penalty than an over-estimation one, allowing for over-subscription buffer. 
From the forecast distribution, we need to pick one of the percentile values to pass onto the \policy PAM module.
The migration counts (not shown) for both very low and high forecast percentiles used for the PAM policy gives suboptimal performance, leading to 1000s of 
migrations; the fewest migrations are for $25\%$'ile and $50\%$'ile.
Low value $<20\%$'ile end up under-estimating the runtime traffic, causing migrations due to constraint violations, while high values $>60\%$ile estimates leads to frequent overestimation of future load and consequently preemptive migrations. 
This is further validated by Figure~\ref{fig:logOfConservativenessViolin}, where one can see the poor performance on both the $P10$ and $P90$ extremes. 
Consistently, the \textit{DRV profiles} (not shown) for $P10$ and $P90$ are worse while $P25$, $P50$ and $P75$ are all better and comparable.
Since ultimately the goal is to pack RUs we chose the $P50$ forecasts as the default in our \policy model.

\section{Related Work}
\label{sec:related}

\textsf{\textit{Benchmark Datasets for Cloud Workloads.}~}
There are no recent public cloud-scale NoSQL workload traces. While there exist~\cite{SurveyCloudDatasets} cloud VM traces~\cite{google52,alibaba6,microsoft85}, Hadoop MapReduce traces~\cite{facebook41,opencloud95} and even DNN/LLM traces~\cite{microsoft104,microsoft16,alibaba7}, real NoSQL database traces are over a decade old~\cite{yahoo130}. Ours is the first open-source NoSQL workload trace, at scale. Our traces operate at the \textit{resource layer} of benchmarking, requiring a simulator such as \simtool to evaluate a placement policy and report resource metrics. In contrast, traditional database benchmarks operate at the \textit{application layer}~\cite{kim2022m2bench,khelifati2023tsm}, with a list of queries as workload and common performance metrics.
Much the research around workload optimization for cloud-hosted databases such as Redis~\cite{RedisWycsb} use Yahoo Cloud Service Benchmarking (YCSB) tool with Application Performance Manager (APM) modules~\cite{APM,YCSB}.
However, most view this as a cloud resource management issue, and assume that generic solutions validated on the above benchmarks apply to large-scale managed cloud databases. \textit{But they do not consider the nuances of distributed, managed and hosted database services, that bring with it subtleties such as redundancies (replicas), sharding, migrations, and faults.}
Some works to attempt to narrow down on load optimization for database services the above, such as ResTune\cite{ResTune} that benchmark with OLTPBench TPC-C \cite{TPC-H} and SYSBENCH.
But they focus on either OLTP or OLAP databases but not both, and the benchmarks only characterize query distributions and transactions, not fault rates, packing densities or loads.

\textsf{\textit{Database Partitioning and Placement.}~}
There are several works that examine placement of database partitions on distributed machines using various quality and performance metrics.
Some try and spread the load across servers for load balancing while others try and co-locate them to avoid distributed execution.
\textit{Zaharia et al.}~\cite{zaharia-vldb-2022} explore the placement of shards on different servers such that queries that access multiple shards can do so in parallel, i.e., shards that are concurrently accessed are placed on different servers to reduce the tail-latency for query execution in OLTP workloads. This is solved using an MILP formulation that decomposes the problem into sub-problems, which can each be solved within seconds.

Others have taken an opposing view, of trying to minimize the number of stores that are accessed to execute a singe query, with the goal of minimizing data movement and coordination~\cite{hermes-vldb-2023}. 
\textit{Accordion}~\cite{accordion-vldb-2014} uses affinity between partitions, i.e., the those that are co-accessed within transactions, to estimate the load on servers hosting the partitions. It dynamically changes the partition placement on servers by solving a MILP optimization problem, to minimize the data movement while ensuring the throughput and memory capacity of a server is not violated and the replication factor is met.

In contrast, in our work, the database partitioning and replication factor have been decided by Cosmos DB. Our goal is limited to placing these partition replicas on the right set of machines, with fewer degrees of freedom. Further, we consider reliability as the primary metric rather than query latency. We also develop a forecasting model to predict and elastically change the placement.

Others take a similar approach as ours to use a forecasting model for query loads and a performance prediction model to decide placement.
\textit{OptimusCloud}~\cite{optimus-atc-2020} does joint tuning of VM configuration and databases for NoSQL databases under dynamic workloads. 
\textit{Wu et al.}~\cite{wu-sigmod-2021} note that past and current load on the system is not a good indicator of future load. They predict the future transaction workload and use this for query routing, load balancing, re-partitioning and live migration. 
While we employ forecasts, we instead use DRV as a novel user-facing metric to optimize for. We leverage our non-parametric simulator 
to evaluate our packing policies.

\textit{Eigen}~\cite{eigen-vldb-2023} uses a hierarchical scheduler for hot, warm and cold servers available to scheduling database instances. It does online resource allocation and offline rebalancing with migration for available machines, using vectorized resource optimizer (VRO), to pack instances into as few servers as possible. It uses warm machines as buffers to handle surges forecasted using exponential smoothing timeseries model. 
We only consider a single type of server and rather than the traditional approach of packing efficiency based on resource usage, we instead target the user-facing DRV metric, which better reflects the experienced reliability and any incidents that are raised. We also use a forecasting model to guide our placement, but also have a policy simulator to validate our proposed policies to offer high confidence before production deployments.

\section{Conclusions}
\label{sec:conclusions}
In this article, we propose open-source benchmarks for NoSQL workloads from real Cosmos DB traces, define a novel DRV user-centric QoS metric, and develop the \simtool framework to evaluate the effect of PAM policies on the SLA for NoSQL workloads. These benchmark components enable evaluation of alternate forecasting and PAM policies for Cosmos DB workloads using the DRV metric, and also simulate the performance for traces from other NoSQL databases and public/private clouds for various policies by \simtool. Further, we propose \policy to solve an optimization problem for replica placement to improve experienced errors by users, which uses the \forecast load forecasting model. We use our benchmarking pipeline to evaluate these, and out-perform the Simulated Annealing model used in Cosmos DB and a worst-fit based solution. \policy offers clear improvements in the load delivered at lower errors, and reduces the resource footprint by $35\%$ without impacting SLAs. These are currently in production in Cosmos DB.Future work includes validation of alternate NoSQL database and cloud traces using this benchmark pipeline, including proactive measures to consider impact of migration by policies, dynamic adjustments to forecasting over-estimates.

\section{Acknowledgments}
We thank students/staff from the DREAM:Lab, IISc, including Pranjal Naman, Mayank Arya, Kautuk Astu, Nikhil Reddy and Srinidhi Ayyagiri, for help with the experiments, plots and reproducibility.

\clearpage

%% ARXIV STYLE
\bibliographystyle{plain}
\bibliography{references}

\end{document}